\documentclass[pra,twocolumn,showpacs,letterpaper,showpacs,superscriptaddress]{revtex4-1}
\usepackage{graphicx,amsmath,amssymb,amsfonts,latexsym,color,dcolumn,bm,epsfig,subfigure}

\usepackage[normalem]{ulem}

\renewcommand{\imath}[0]{\mathrm{i}}
\newcommand{\abs}[1]{\left\vert#1\right\vert}

\begin{document}

\title{Non-Markovianity in atom-surface dispersion forces}

\author{F. Intravaia}
\affiliation{Max-Born-Institut, 
             12489 Berlin, Germany}
\author{R. O. Behunin}
\affiliation{Department of Applied Physics, Yale University, 
             New Haven, Connecticut 06511, USA}
\author{C. Henkel}
\affiliation{Institute of Physics and Astronomy, University of Potsdam, 
             Karl-Liebknecht-Str. 24/25, 14476 Potsdam, Germany}
\author{K. Busch}
\affiliation{Max-Born-Institut, 
             12489 Berlin, Germany}
\affiliation{Humboldt-Universit\"at zu Berlin, Institut f\"ur Physik, 
             AG Theoretische Optik \& Photonik, 12489 Berlin, Germany}
\author{D. A. R. Dalvit}
\affiliation{Theoretical Division, MS B213, Los Alamos National Laboratory, 
             Los Alamos, NM 87545, USA}

\newcommand{\mathbfh}[1]{\hat{\mathbf{#1}}}

\begin{abstract} 
We discuss the failure of the Markov approximation in the description of  atom-surface fluctuation-induced interactions, both at equilibrium (Casimir-Polder forces) and out-of-equilibrium (quantum friction). Using general theoretical arguments, we show that the Markov approximation can lead to erroneous predictions of such phenomena with regard to both strength and functional dependencies on system parameters. Our findings highlight the importance of non-Markovian effects in dispersion interactions.
In particular, we show that the long-time power-law tails of temporal correlations, and the corresponding low-frequency behavior, of two-time dipole correlations, neglected in the Markovian limit, dramatically affect the prediction of the force.
\end{abstract}

\maketitle


\section{Introduction} 

Prototype examples of dispersion interactions mediated by the quantum electromagnetic field are the Casimir-Polder force on a static atom close to a surface \cite{Casimir48a}, and the quantum frictional force experienced by the same atom as soon as it starts moving above the surface \cite{Pendry97,Volokitin07}. From a theoretical standpoint, one must often rely on approximations to model such interactions and predict the outcome of a specific experimental setup. 
One of these is the Markov approximation, which is also one of the most ubiquitous and successful approximations used in quantum optics,
and has provided reliable predictions for numerous experimental setups. 
In Casimir physics, the Markov approximation is often employed to solve the coupled dynamics of the atom and surface interactions mediated by electromagnetic quantum fluctuations.
The key assumption underlying this approximation is that the system's
memory can be ignored, i.e., the future of its dynamics is only related to the  immediate present. More specifically, the logic behind this approximation finds its justification in the fact that sub-systems often exhibit very different correlation times,  so that on average the fastest dynamics blurs the evolution of the slower sub-system, effectively erasing its memory \cite{Gardiner91,Breuer02}. 

In this work, we argue that the Markov approximation fails to provide reliable 
predictions for non-equilibrium fluctuation-induced interactions. Surprisingly, depending on 
the targeted level of accuracy, it can already lead to incorrect results 
for systems in equilibrium. Its failure becomes, however, more relevant as soon as non-equilibrium 
systems are being considered. 
Although in some circumstances non-Markovian effects are already known to strongly affect
the dynamics of quantum systems (e.g. non-exponential decay of excited quantum states for atoms 
in photonic crystals \cite{John90,Vats02,Hoeppe12}), their impact on fluctuation-induced phenomena 
has not been explored.
Moreover, the role and relevance of non-Markovian
effects in conjunction with non-equilibrium physics, has yet to be thoroughly investigated.
Our discussion focuses on the prototypical system consisting of a single atom (or, in general, 
a microscopic system with internal degrees of freedom) interacting with a planar surface (see
Figs. \ref{CP} and \ref{friction}). 
Besides the large interest of modern experiments in such setups 
\cite{Intravaia11},
e.g., in thermal radiation and hybrid atom-chip systems,
they lend themselves to advanced but still not too complex theoretical treatments.

Our paper is organized as follows. In Section \ref {static} we review the 
theory of equilibrium atom-surface interactions and derive the expression for the Casimir-Polder 
force. The purpose of this part is twofold: On the one hand, it allows for an exposition of
the formalism and the basic concepts which will be used in the non-equilibrium case.
On the other hand, our approach allows for considerations beyond standard perturbative techniques 
\cite{Buhmann04,Intravaia11} 
and the usual Lifshitz theory of equilibrium phenomena
\cite{Dzyaloshinskii61}. 
In particular, we show that long-time tails in the two-time dipole correlation (and its corresponding low-frequency behavior) arising from non-Markovian effects  become relevant as soon as one goes beyond  second-order perturbation theory (see Fig. \ref{PowerSpectrumComp}). Since Casimir interactions are a broad-band frequency
phenomenon, a precise description of the behavior at low frequencies is fundamental for the evaluation of the interaction.
Our final result will formally take into account the general
response of the atomic system, and the standard Lifshitz formula is obtained 
as a special case in the linear approximation.

Section \ref{dynamical}  focuses on quantum friction.
Like ordinary friction, quantum friction describes a force acting on an object moving near another one (see Fig. \ref{friction}).
Unlike the classical case, however, quantum friction is mediated by the interaction with the quantum electromagnetic field at zero temperature.
We argue that, as in the static case, non-Markovian effects affecting the low-frequency behavior of the dipole correlator are crucial for the correct evaluation of this drag force (see Eq. \eqref{quantity} below). In fact,
these effects are responsible for the velocity-dependence of the frictional force. As we shall demonstrate in this work,
it is precisely this sensitivity of quantum friction to the
Markov approximation that explains the vast zoology of very different results found in the literature, e.g. 
\cite{Volokitin07,Scheel09,Barton10b,Intravaia14,Intravaia15}.

Finally, in Section \ref{2ndOrderParagraph} we discuss quantum friction within lowest-order perturbation theory, and analyze the impact of intrinsic or induced dissipation on the drag force.


\section{Non-Markovian effects within the Casimir-Polder interaction} 
\label{static}

\begin{figure}[t!]
\includegraphics[width=7.5cm]{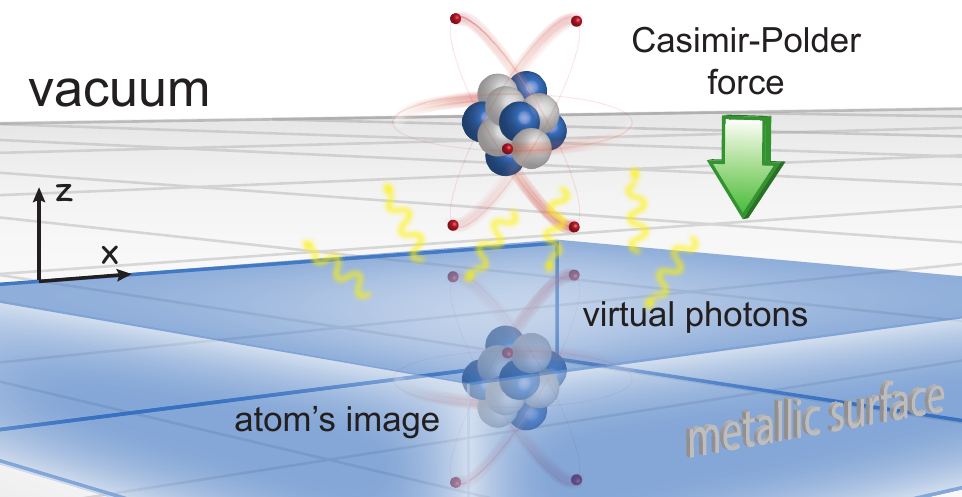}
\vspace{-.3cm}
\caption{Schematic of an atom above a surface experiencing the Casimir-Polder force. The presence
of the correlation tensor in the expression of the force can be qualitatively understood as stemming from the interaction of the atom with its image within the material.
}
\label{CP}
\end{figure}

Let us start by considering the familiar Casimir-Polder interaction between an atom located 
at position $\mathbf{r}_{a}$ above a semi-infinite planar medium (see Fig. \ref{CP}). 
Within the open quantum system framework, the resulting dissipation can be thought of as 
stemming from the interaction of the matter's degrees of freedom with a dissipative bath 
\cite{Huttner92,Rosa10,Rosa11,Intravaia12b}. 
In our description, the atom is represented by the dipole operator $\mathbfh{d}(t)$. For symmetry 
reasons, the force acting on the atom has only a single component along the direction normal to 
the planar interface (which we denote as the $z$-direction). At any given time $t$, it is given 
by
\begin{equation}
 F_{\rm CP}(t)=\langle\mathbfh{d}(t)\cdot\partial_{z_a}\mathbfh{E}(\mathbf{r}_{a},t)\rangle ,
\label{Fcp}
\end{equation}
where $\mathbfh{E}(\mathbf{r},t)$ is the electric field operator,
${\bf r}_a$ is the position of the atom,  and the expectation value
is taken over the initial state of the system.
This expression for the atom-surface force can be derived from the Lorentz force on the atom. 
\cite{Buhmann04}. 
It is important to stress that the time dependence of the operators in Eq.\eqref{Fcp} is the 
full time evolution dictated by the total system and, therefore, it includes the effect of the 
interactions between the different subsystems (atom, field, and matter). 
%
Using the Maxwell equations, the electric field operator can be written as 
$\mathbfh{E}(\mathbf{r},t)=\mathbfh{E}^{(+)}(\mathbf{r},t)+{\rm h.c.}$, where
\begin{multline}
\mathbfh{E}^{(+)}(\mathbf{r},t)=\mathbfh{E}_{0}^{(+)}(\mathbf{r},t)\\
+\frac{\imath}{\pi}\int_{0}^{\infty}d\omega\, \int_{t_{i}}^{t} dt' e^{-\imath \omega(t-t')}\underline{G}_{I}(\mathbf{r},\mathbf{r}_{a},\omega)\cdot\mathbfh{d}(t')
\label{emfield}
\end{multline}
with $t_{i}$ being some initial time. Here, $\mathbfh{E}_{0}^{(+)}$ is the positive-frequency 
part (related to the annihilation operators) of the electromagnetic field in the absence of 
the atom but in the presence of the dissipative medium, and
$\underline{G}(\mathbf{r},\mathbf{r}', \omega)$ is the Green tensor of the half-space associated
with the surface. To derive equation \eqref{emfield} we used the Kramers-Kronig relations for the Green tensor (see Appendix \ref{MaxwellG}). 
Hereafter we adopt the subscripts ``$I$'' and ``$R$'' to, respectively, indicate  
the imaginary and the real part of a quantity (component-wise for tensorial objects). Because the dipole and the field (positive- 
and negative-frequency parts separately) operators commute at equal times, by normal ordering 
we can write 
%
\begin{align}
F_{\rm CP}(t)&=\langle\mathbfh{d}(t)\cdot\partial_{z}\mathbfh{E}^{(+)}(\mathbf{r}_{a},t)\rangle+{\rm h.c.}\nonumber\\
&=2\mathrm{Re}\langle\mathbfh{d}(t)\cdot\partial_{z}\mathbfh{E}^{(+)}(\mathbf{r}_{a},t)\rangle .
\label{EReal}
\end{align}
The final result cannot depend on the ordering we choose, as long as this ordering
is consistently used throughout the entire derivation. However, as will  appear in the 
following, this specific choice of ordering is convenient for our calculations. 
Physically it has the implication of attributing the force exclusively to the radiation reaction 
in the atom dynamics 
\cite{Milonni73}, 
while the contribution of the vacuum field related to the operator $\mathbfh{E}_{0}$ seemingly
disappears. However, this does not mean that the quantum properties $\mathbfh{E}_{0}$ are 
irrelevant: In this approach they are ``hidden" in the expression for the time-dependent dipole operator 
and they will explicitly appear again when we consider, for example, dipole correlation functions.
For other choices, such as symmetric ordering 
\cite{Dalibard82},
both the dipole and the field itself contribute to the force and the term containing 
$\mathbfh{E}_{0}$ must consistently be kept throughout the calculation.

To evaluate the force, let us assume an initially factorized state 
$\hat{\rho}(t_{i})=\hat{\rho}_a(t_{i})\bigotimes\hat{\rho}_{\rm f/m}(t_{i})$, where 
$\hat{\rho}_a(t_{i})$ is the atom's initial density matrix and $\hat{\rho}_{\rm f/m}(t_{i})$ 
represents the state of the coupled field/matter system. Both subsystems are assumed to be 
initially in their respective ground states, which implies that 
$\langle\mathbfh{d}(t)\cdot\partial_{z_a}\mathbfh{E}_{0}^{(+)}(\mathbf{r}_{a},t)\rangle
 =
 \mathrm{tr}[\mathbfh{d}(t)\cdot\partial_{z_a}\mathbfh{E}_{0}^{(+)}(\mathbf{r}_{a},t)\hat{\rho}(t_{i})]=0$  
for all times (here, the symbol ``$\mathrm{tr}$'' traces over the quantum states). This means 
that the contribution coming from the vacuum state of field plus medium subsystem drops out of the 
calculation. Note that, although both subsystems are initially in their respective ground states, 
in general, the state $\hat{\rho}(t_{i})$ \emph{is not} the ground state of the composite system 
\cite{Ford06,Intravaia08}.
During the course of the time-evolution, the atom subsystem becomes entangled with the field/matter 
subsystem, undergoing the well-known ``dressing" process \cite{Compagno05}.

Using the decomposition in Eq. \eqref{emfield} 
and the symmetry properties of the Green tensor, 
$G_{ij}(\mathbf{r},\mathbf{r}',\omega)=G_{ji}(\mathbf{r}',\mathbf{r},\omega)$, we can rewrite the 
atom-surface force as 
\begin{multline}
F_{\rm CP}(t)=
\mathrm{Re}\bigg(\frac{2\imath}{\pi}\int_{0}^{\infty}d\omega\, \int_{0}^{t-t_{i}}d\tau e^{-\imath \omega \tau}\\
\times\mathrm{Tr}\left[\underline{C}(t,t-\tau)\cdot \partial_{z}\underline{G}_{I}(\mathbf{r}_{a},\mathbf{r},\omega)_{\vert \mathbf{r}=\mathbf{r}_{a}}\right]\bigg) ,
\label{source1}
\end{multline} 
where ``$\mathrm{Tr}$'' traces ove tensor indices and we have set $\tau=t-t'>0$. 
The tensor $\underline{C}$ is the (non-symmetrically ordered) two-time dipole correlator,
\begin{equation}
C_{ij}(t,t-\tau)=\langle \hat{d}_{i}(t) \hat{d}_{j}(t-\tau)\rangle ,
\label{staticCorrelator}
\end{equation} 
which plays a key role in what follows. The coupled equations of motion for the atom and 
field/matter can be solved (e.g. numerically or using the Born-Markov approximation \cite{Buhmann04})
to obtain the dynamic evolution of the two-time dipole correlator and the time-dependent atom-surface force. 

A natural question that arises is whether non-Markovian effects are relevant in the dynamics 
of the dipole correlator and, consequently, in the force $F_{\rm CP}(t)$. 
In order to address this question, we will consider the limit of large times in which the entire 
system of atom plus field/matter equilibrates and evolves to a stationary state. 
In this case, the correlator depends only on the time difference $\tau$, i.e.,
$C_{ij}(\tau)= {\rm tr}[ \hat{d}_i(\tau) \hat{d}_j(0) \hat{\rho}(\infty)]$, 
where $\hat{\rho}(\infty)$ represents the stationary density matrix of the full system. In 
this limit the atom-surface force reaches the constant value 
$F_{\rm CP} \equiv \lim_{t \rightarrow \infty} F_{\rm CP}(t)$, which should coincide with the 
expression of the well-known Casimir-Polder interaction \cite{Casimir48a}. 
In order to evaluate Eq.(\ref{source1}) for large times, it is convenient to introduce 
the dipole power spectrum tensor $\underline{S}(\omega)$, which is defined in terms 
of the correlator as
\begin{equation}
\underline{S}(\omega ) =  \frac{1}{2\pi}\int_{-\infty}^{\infty}d\tau e^{\imath\omega \tau} \underline{C}(\tau) .
\end{equation}
As  $\underline{C}^{\dagger}(\tau)=\underline{C}(-\tau)$, the power spectrum is Hermitian 
$\underline{S}^{\dagger}(\omega)=\underline{S}(\omega)$ and its real part $\underline{S}_{R}(\omega)$ 
is a symmetric matrix, while its imaginary part $\underline{S}_{I}(\omega)$ is anti-symmetric. 
Using these properties, the stationary Casimir-Polder force can be written as
\begin{multline}
F_{\rm CP}=
\frac{2}{\pi}\int_{0}^{\infty}d\omega\, \int_{-\infty}^{\infty}d\nu \\
\times \mathrm{P}\left(\frac{\mathrm{Tr} \left[\underline{S}_{R}(\nu)\cdot \partial_{z}\underline{G}_{I}(\mathbf{r}_{a},\mathbf{r},\omega)_{\vert \mathbf{r}=\mathbf{r}_{a}}\right]}{\omega+\nu}\right)~,
\label{source1bis}
\end{multline} 
where $\mathrm{P}$ denotes the principal value and we have used that the matrix 
$\partial_{z}\underline{G}_{I}(\mathbf{r}_{a},\mathbf{r}, \omega)_{\vert \mathbf{r}=\mathbf{r}_{a}}$ 
is symmetric (see Appendix \ref{MaxwellG}), and trace-orthogonal to any anti-symmetric matrix.

As we will show below, depending on the approach and the approximations used to compute 
this stationary dipole-dipole power spectrum, one obtains different expressions for the 
stationary interaction.


\subsection{Using the fluctuation-dissipation theorem}
\label{StaticCPFDT}

At large times one can invoke equilibrium considerations to determine the stationary density 
matrix. We will assume that the full system thermally equilibrates at an inverse temperature 
$\beta$, and that the stationary state is given by a Gibbs state 
$\hat{\rho}(\infty) \propto e^{-\beta \hat{H}}$ 
where 
$\hat{H}$ describes the Hamiltonian of the full system 
\cite{Kubo57,Kubo66,Haag67,Breuer02}, 
including the couplings between all sub-systems.
In these circumstances, we can appeal to the fluctuation-dissipation theorem (FDT) 
\cite{Callen51}. 
This theorem of equilibrium thermodynamics establishes a connection between the power spectrum 
and the linear response of the system to a small external perturbation. In particular, we will 
consider the case of zero temperature ($T = 0$), for which the Gibbs state is the highly 
entangled ground state of the full system.
In this case, for a non-symmetrized correlator, the FDT takes the form
\cite{Kubo66}
\begin{equation}
 \underline{S}(\omega)=\frac{\hbar}{\pi}\theta(\omega)\underline{\alpha}_{\Im}(\omega),
\label{FDT_general}
\end{equation}
where $\theta(\omega)$ denotes the Heaviside function, and
$\underline{\alpha}_{\Im}(\omega) =  [\underline{\alpha}(\omega)-\underline{\alpha}^{\dag}(\omega)]/ (2\imath)
=\underline{\alpha}_{I}^{\rm s}(\omega)-\imath\underline{\alpha}_{R}^{\rm as}(\omega)$ (the superscripts ``${\rm s}$'' and ``${\rm as}$'' indicate the symmetric and the anti-symmetric part of the tensor). 
Further, $\underline{\alpha}(\omega)$ represents the atom's 
complex susceptibility (polarizability) tensor, i.e. the Fourier transform of Kubo's formula 
in the case of linear response, i.e.,
\begin{equation}
{\alpha}_{ij}(\tau) = \frac{\imath}{\hbar} \theta(\tau)\mathrm{tr} \{ [\hat{d}_i(\tau),\hat{d}_j(0)]\hat{\rho}(\infty) \} .
\end{equation}
As any susceptibility (including the Green 
tensor), $\underline{\alpha}(\omega)$ is analytic in the upper part of the complex $\omega$-plane
and satisfies the crossing relation $\underline{\alpha}(-\omega^{*})=\underline{\alpha}^{*}(\omega)$. 
It follows that the real (imaginary) part of the polarizability is an even (odd) function of 
frequency. Using these properties in combination with the FDT and the Kramers-Kronig relations, 
we may rewrite Eq.\eqref{source1bis} as \cite{Wylie84}
\begin{equation}
F_{\rm CP} =\frac{\hbar}{\pi}
\int_{0}^{\infty}d\xi \, \mathrm{Tr}\left[\underline{\alpha}^{\rm s}(\imath \xi,\mathbf{r}_{a})\cdot \partial_{z}\underline{G}(\mathbf{r}_{a},\mathbf{r}, \imath \xi)_{\vert \mathbf{r}=\mathbf{r}_{a}}\right]~.
\label{imgFDT}
\end{equation}
Here, we exploited the analytic properties of $\underline{\alpha}(\omega,{\bf r})$ and $\underline{G}({\bf r}, {\bf r'},\omega)$ 
in the complex $\omega$  plane to express the final result as an integral along the positive imaginary-frequency axis (Wick rotation). 
In the above expression, we have explicitly indicated the position dependence of the atom's 
polarizability to stress that the dressing is depending on the system's geometry. Although it 
is redundant because of the properties of the Green tensor, we have also used the superscript 
``${\rm s}$'' to emphasize that only the symmetric part of the polarizability is relevant for 
the final result.

As a final remark of this subsection, we want to emphasize that, despite certain formal similarities, 
Eq.\eqref{imgFDT} differs from standard formulas found in the literature for the atom-surface interaction.
Indeed, Eq.\eqref{imgFDT} contains the exact atomic polarizability, while standard expressions are special cases
of this formula, e.g. the Lifshitz formula for linear systems \cite{Dzyaloshinskii61}, and 
perturbative expansions in the atom-field coupling strength \cite{Casimir48a,Wylie84,McLachlan63}.
For further details, see Appendix \ref{GenericCP}. 


\subsection{Using the quantum regression theorem}

In quantum optics, one of the most widely used tools to evaluate two-time correlators
is the quantum regression theorem (QRT) 
\cite{Lax63,Mandel95,Buhmann04}. 
Often considered as the quantum extension of Onsager's regression conjecture 
\cite{Onsager31,Onsager31a}, 
the QRT finds its justification within the framework of master equations. Using the Born 
approximation (which entails the factorization of  the density matrix at any time, i.e., no 
back-action of the system onto the environment),
it is possible to show that the regression of fluctuations is identical to the decay of a 
one-time average
\cite{Lax63}. 
In other words, the QRT shows that the equations of motion for the correlations are the same of those for the mean values. 
Very often, however, these equations of motion can only be solved by using the Markov approximation (no memory effects), 
thus requiring the validity of the Born-Markov approximation for applying the QRT.
Although the QRT provides expressions for correlators 
that are more explicit than those in the FDT, the approximations on which it relies limit its 
validity to the limit of weak atom-field coupling and to a narrow range of frequencies close 
to a resonance 
\cite{Talkner86,Ford96a,Ford00,Lax00,Ford00a}. 

To show how this affects the Casimir-Polder interaction, we now evaluate the two-time dipole 
correlation tensor introduced at the beginning of this section using the QRT approach. Let us  
model the atom as a multi-level system, and consider the operators 
$\hat{A}_{mn}=|m\rangle\langle n|$ that describe the transition from the atomic level 
$|n\rangle$ to $|m\rangle$. 
We write the dipole operator as $\mathbfh{d}(t)=\mathbf{d}\hat{q}(t)$, where $\mathbf{d}$ denotes
the (real) dipole vector and $\hat{q}(t)$ is a dimensionless operator describing the quantum 
dynamics of the internal degrees of freedom. 
For this operator we have $\hat{q}(t)=\sum_{nm} c_{nm}\hat{A}_{mn}(t)$, where $c_{nm}$ are the 
matrix elements corresponding to the transition described by $\hat{A}_{mn}$. According to the 
QRT, the  two-time correlator $\langle A_{nm}(t) A_{m'n'}(t') \rangle$ is given by 
\cite{Mandel95,Breuer02,Buhmann04} ($t>t'$)
\begin{multline}
\langle A_{mn}(t)A_{m'n'}(t')\rangle=\delta_{nm'}\langle A_{mn'}(t')\rangle \\
\times e^{[\imath\tilde{\omega}_{mn}(\mathbf{r}_{a})-\gamma_{mn}(\mathbf{r}_{a})](t-t')} ,
\label{QRTtrans}
\end{multline}
where $\tilde{\omega}_{mn}(\mathbf{r}_{a})$ and $\gamma_{mn}(\mathbf{r}_{a})>0$ respectively
represent the position-dependent, field-renormalized transition frequencies and decay rates 
corresponding to the $n\to m$ transitions. In the usual master equation framework it is expected 
that, in the absence of any external driving force and at large times, the reduced density matrix 
of the atom should relax to  
$\hat{\rho}_a(\infty) \simeq e^{-\beta \hat{H}_a}$
\cite{Breuer02,Buhmann04},
where $\hat{H}_a$ is the Hamiltonian of the atom subsystem (note that this state is different from the one used in the FDT above).
In particular, at zero 
temperature this reduced density matrix should relax to the atomic ground state 
\cite{Note1}.
Therefore, $\langle A_{mn}(\infty)\rangle=\delta_{m0}\delta_{n0}$ and
$\lim_{t\rightarrow \infty} \langle A_{mn}(t)A_{m'n'}(t-\tau)\rangle 
 =
 \delta_{nm'} \delta_{m0} \delta_{n'0} \exp ([\imath\tilde{\omega}_{0n}(\mathbf{r}_{a})-\gamma_{0n}(\mathbf{r}_{a})]\tau) $ 
\cite{Buhmann04}, 
featuring an exponential decay of correlations. In the simple case of a two-level system, the stationary dipole correlator is then given by  
\cite{Mandel95,Breuer02,Buhmann04} ($\tau>0$)
\begin{align}
C_{ij}(\tau)=d_i d_j \,e^{\left[-\imath\tilde{\omega}_{a}(\mathbf{r}_{a})-\gamma_{a}(\mathbf{r}_a)\right]\tau} ,
\label{QRT-TLS}
\end{align}
where  $\tilde{\omega}_{a}=\tilde{\omega}_{10}$ and $\gamma_{a}=\gamma_{01}$.
This implies that the stationary limit of Eq. \eqref{source1} is given by
\begin{align}
F^{\rm QRT}_{\rm CP} =
\mathrm{Re}\left[\frac{2}{\pi}\int_{0}^{\infty}d\omega\,  \frac{\mathrm{Tr}\left[\mathbf{d}\mathbf{d}\cdot \partial_{z}\underline{G}_{I}(\mathbf{r}_{a},\mathbf{r}, \omega)_{\vert \mathbf{r}=\mathbf{r}_{a}}\right]}{\omega+\tilde{\omega}_{a}(\mathbf{r}_{a})-\imath\gamma_a(\mathbf{r}_{a})}\right] .
\label{sourceB}
\end{align} 
Using the symmetry properties of the Green tensor, we can rewrite Eq. \eqref{sourceB} as an integral 
along the imaginary axis  (see Appendix C)%
\begin{multline}
F^{\rm QRT}_{\rm CP}
=\frac{\hbar}{\pi}\int_{0}^{\infty}d\xi \, \mathrm{Tr}\bigg{[}\left(\frac{\underline{\alpha}^{(2)}(\imath\xi, \mathbf{r}_{a})+\underline{\alpha}^{(2)}(-\imath\xi, \mathbf{r}_{a})}{2}\right)\\
\times \partial_{z}\underline{G}(\mathbf{r}_{a},\mathbf{r}, \imath\xi)_{\vert \mathbf{r}=\mathbf{r}_{a}}\bigg{]} ,
\label{imgQRT}
\end{multline} 
where 
\begin{equation}
\underline{\alpha}^{(2)}(\omega, \mathbf{r}_{a})=\frac{\mathbf{d}\mathbf{d}}{\hbar}\frac{2 \tilde{\omega}_{a}(\mathbf{r}_{a}) }{\tilde{\omega}^{2}_{a}(\mathbf{r}_{a})-(\omega+\imath\gamma_a(\mathbf{r}_{a}))^{2}}
\label{alpha2ndOrder}
\end{equation}
denotes the standard atomic polarizability tensor for a two-level atom computed in second-order perturbation 
theory \cite{Milonni04,Buhmann04,Lach12,Jentschura15}. To some extent, $\underline{\alpha}^{(2)}(\omega, \mathbf{r}_{a})$ can be regarded as a generalization 
to higher orders of perturbation theory of the corresponding bare polarizability 
$\underline{\alpha}^{(0)}(\omega) 
 = 
 (2\omega_{a}/\hbar)\mathbf{d}\mathbf{d}\left[\omega_{a}^{2}-(\omega+\imath 0^{+})^{2}\right]^{-1}$ 
(the small imaginary part in the denominator is introduced in order to enforce causality).
Equation (\ref{imgQRT}) was first derived in  
\cite{Buhmann04} 
using the QRT approach.

\begin{figure}[t]
\includegraphics[width=8.5cm]{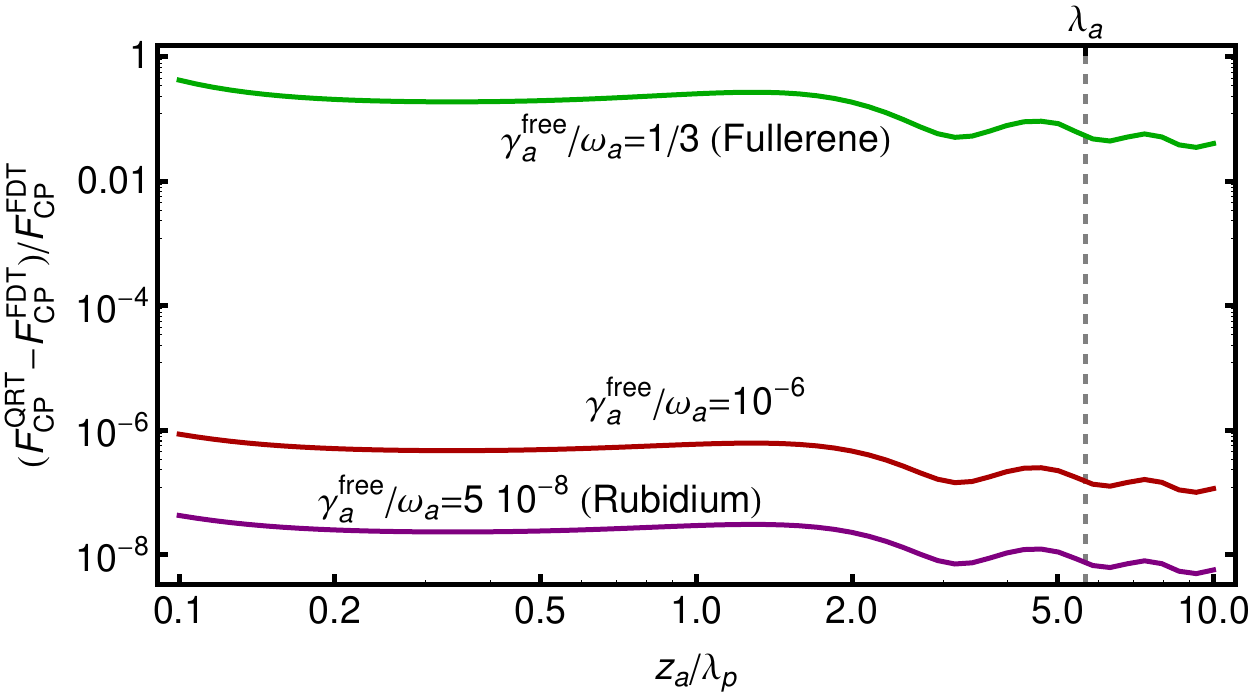}
\caption{
	Non-Markovian correction to the Casimir-Polder force as a function of the atom-surface separation. 
	The three curves correspond to different values of the atom-field coupling: 
	The lower (purple) curve refers to $^{87}$Rb (static polarizability 
	$\alpha_0=2\abs{\mathbf{d}}^{2}/{3\hbar\omega_{a}}=5.26 \times 10^{-39} {\rm F} {\rm m}^2$  
	\cite{Steck08}); 
	the middle (red) curve corresponds to a hypothetical two-level system with free decay rate, 
	$\gamma^{\rm free}_{a}/\omega_{a}= 10^{-6}$, where $\omega_{a}$ is the transition frequency of $^{87}$Rb; 
	the upper (green) curve is the result for $\gamma^{\rm free}_{a}/\omega_{a}=1/3$, typical of fullerene 
	(C$_{60}$, see Ref. 
	\cite{Buhmann12}). 
	For simplicity, all three cases have the same value for the transition frequency. The 
	oscillations visible at large separations are associated with the oscillations of the decay rates for 
	distances $z_{a}\lambda_{a}\gtrsim1$. The planar surface is modeled as a metallic half-space
	whose constituent material is described by the Drude model 
	$\epsilon(\omega)=1-\omega_{p}^{2}[\omega(\omega+\imath \Gamma)]^{-1}$ 
	with parameters that are typical for gold: $\omega_{p}=9$ eV and $\Gamma/\omega_{p}= 5 \times 10^{-3}$.
	The distance is measured in units of the plasma wavelength $\lambda_{p}=2\pi c/\omega_{p}$ ($\sim 140 $ nm).
}
\label{CasimirPolderNM}
\end{figure}

Upon comparing the expression for the Casimir-Polder force obtained with the QRT, Eq. \eqref{imgQRT}, with 
that obtained using the FDT, Eq. \eqref{imgFDT}, we note that they only coincide at the lowest order in 
perturbation theory, or, equivalently, when the polarizabilities in both equations are replaced by the bare 
polarizability. At higher orders they clearly differ because of the drastically distinct methods used to 
calculate the stationary two-point dipole correlation tensor $\underline{C}(\tau)$. The first and most 
evident difference is in the dependence of $F_{\rm CP}$ on the spectral line-width of the polarizability. 
Compared to the result in Eq. \eqref{imgFDT}, with $\underline{\alpha}(\omega, \mathbf{r}_{a})\approx \underline{\alpha}^{(2)}(\omega, \mathbf{r}_{a})$, 
the force resulting from Eq. \eqref{imgQRT} is less sensitive to the radiative 
decay rate and its magnitude only slightly deviates from the result obtained using the bare polarizability. The application of the 
QRT results in a force whose magnitude is larger than that obtained via the FDT, 
and the difference is more pronounced for atoms with larger dipole moment (stronger atom-field coupling) 
\cite{Buhmann12}. 
We depict this behavior in Fig \ref{CasimirPolderNM}, where we compare the FDT- and QRT-based predictions 
using the polarizability in Eq. \eqref{alpha2ndOrder} (for simplicity we neglected the surface-induced 
frequency shift).

\subsection{Relation between the FDT and the QRT}
\label{FDTvsQRT}

The result of the previous two sub-sections demonstrate that the use of two of the most popular 
approaches for calculating quantum correlators leads to different results for the Casimir-Polder 
force beyond leading order.
It is important to emphasize that 
while the FDT is an exact theorem, the QRT represents only an approximation. 
It has already been pointed out that, because of the Born-Markov approximation, the QRT may lead to results 
that are incompatible with the statistical mechanics of quantum systems when treated beyond the weak-coupling 
approximation or perturbed far away from resonance 
\cite{Talkner86,Ford96a}. 
The quantum-regression theorem has proven to be remarkably successful for driven quantum optical 
systems, where its range of validity, near resonance, is not an impediment \cite{Ford00,Lax00,Ford00a}. 
However, because of 
the broadband nature of electromagnetic fluctuation-induced interactions, this limited range of validity makes the 
QRT in general inadequate to treat such interactions. 
An inappropriate description of any part of the spectrum can lead to erroneous results. 

Recently, in an attempt to quantify the dynamical properties of an open quantum system,
the failure of the QRT has been proposed as a measure of the system's degree of non-Markovianity \cite{Gullo14,Ali15}.
In fact, the exponential behavior in the correlator obtained by applying the QRT is a direct consequence
of the Markov approximation. It is well known, however, that this behavior 
is incorrect at large times, where the exponential decay of the correlations transforms to power-law decay \cite{Knight76,Berman10}. 
This difference has been investigated in other contexts of quantum optics, e.g., the 
dynamics of the quantum harmonic oscillator 
\cite{Ford96a} 
or the spontaneous decay of an excited atom in the electromagnetic vacuum  
\cite{Davidson70,Davidson71,Wodkiewicz76,Knight76}. 
This phenomenon is related to the limitations of the Wigner-Weisskopf approximation \cite{Weisskopf30,Foerster72}, which,
in turn, is  equivalent to the Markov approximation \cite{Knight76,Berman10}.
A similar argument applies to high frequencies, since it is also 
known that for short times the decay process starts quadratically in $\tau$ instead of the linear behavior 
associated with an exponential law \cite{Cohen-Tannoudji98}. Consequently, when one goes from the time to the frequency domain, the Fourier transform of the correlator obtained using the QRT becomes imprecise at low and high frequencies \cite{Ford96a}.

To further understand how non-Markovian effects are responsible for the difference between the two 
expressions for the Casimir-Polder force, it is convenient to examine with some detail the two-time correlator $\underline{C}(\tau)$
in the limit of large times. According to Eq. \eqref{source1} and the FDT, the relevant part of this 
quantity for $F_{\rm CP}$ is given by
\begin{multline}
\underline{C}^{\rm s}(\tau)=\frac{\hbar}{\pi}\int_{0}^{\infty} d\omega e^{-\imath\omega\tau}\underline{\alpha}^{\rm s}_I(\omega)
 \\
 =-\hbar \sum_{i} \! \mathrm{Res}[\underline{\alpha}^{\rm s}(\Omega_{i})]e^{-\imath\Omega_{i}\tau} \\
-\!\imath\frac{\hbar}{\pi} \!\int_{0}^{\infty} \!\!d\xi  
e^{-\xi\tau}[\underline{\alpha}^{\rm s}_{I}(\omega)]_{\vert\omega=-\imath\xi+0^{+}}~, 
\label{exactC}
\end{multline}
where, for simplicity, we have suppressed the dependence of the polarizability on $\mathbf{r}_{a}$. The 
second line is obtained by computing the $\omega$ integral using a contour in the lower right quadrant 
of the complex frequency plane. Here,  ``Res'' denotes the residue, and  $\Omega_{i}=\Omega_{i}(\mathbf{r}_{a})$ 
are the complex poles of the polarizability. 
As these poles are located in the lower right quadrant, we can write  $\Omega_{i} =  \omega_{i}(\mathbf{r}_{a})- \imath\gamma_{i}(\mathbf{r}_{a})$,
where $\omega_{i}(\mathbf{r}_{a})$ and $\gamma_{i}(\mathbf{r}_{a})$ are two non-specified real and positive 
functions whose exact expression depends on the specific model for the atom (e.g. for the two-level atom they are given in Eq.
(\ref{QRT-TLS})). At this point, we would like
to note that, for simplicity, we have assumed that the polarizability has 
no other discontinuity but isolated poles in 
the complex plane. If this were not the case, e.g., when branch cuts would be present, they must be added 
to Eq.(\ref{exactC}). For the present case of simple poles, we see that the stationary dipole correlation, 
as given by the FDT, contains a decaying exponential behavior just like the QRT (first term in Eq.(\ref{exactC})), 
\emph{plus} an extra term which is ultimately responsible for the difference between Eqs. \eqref{imgFDT} and 
\eqref{imgQRT}. Upon an analytical continuation, one has 
\begin{equation}
[\underline{\alpha}^{\rm s}_{I}(\omega)]_{\vert\omega=-\imath\xi+0^{+}} = -  
\frac{ \underline{\alpha}^{\rm s}(\imath\xi)-\underline{\alpha}^{\rm s}(-\imath\xi)}{2\imath}.
\end{equation}
This expression yields the terms that are missing in Eq.\eqref{imgQRT} in order to recover Eq.\eqref{imgFDT} 
(see Appendix \ref{ComplexFreq}). While the exponential terms $\exp(-\imath \Omega_i \tau)$ in Eq.\eqref{exactC}  
are sensitive to the details of the polarizability for frequencies around the poles $\Omega_{i}$, the last 
integral in Eq.\eqref{exactC} when $\tau \to \infty$ is sensitive to the behavior of the polarizability at 
low frequencies, $|\omega|\sim 0$. From the crossing relation, we can deduce that $\underline{\alpha}^{\rm s}_{I}(\omega)$ 
is odd in frequency, which means that its form is $\underline{\alpha}^{\rm s}_{I}(\omega) \approx a_{2m+1} \omega^{2m+1}$ 
($m=0,1,2,\dots$). As a result, for times $\tau$ for which the exponentially decaying terms have died out, i.e., 
when $\tau \gg 1/{\rm min}(\gamma_i)$, the stationary correlation behaves as
\begin{equation}
\underline{C}^{\rm s}(\tau)\approx  \frac{\hbar}{\pi} a_{2m+1}(-1)^{m+1}\frac{(2m+1)!}{\tau^{2(m+1)}} .
\end{equation}

\begin{figure}[t]
\includegraphics[width=8.5cm]{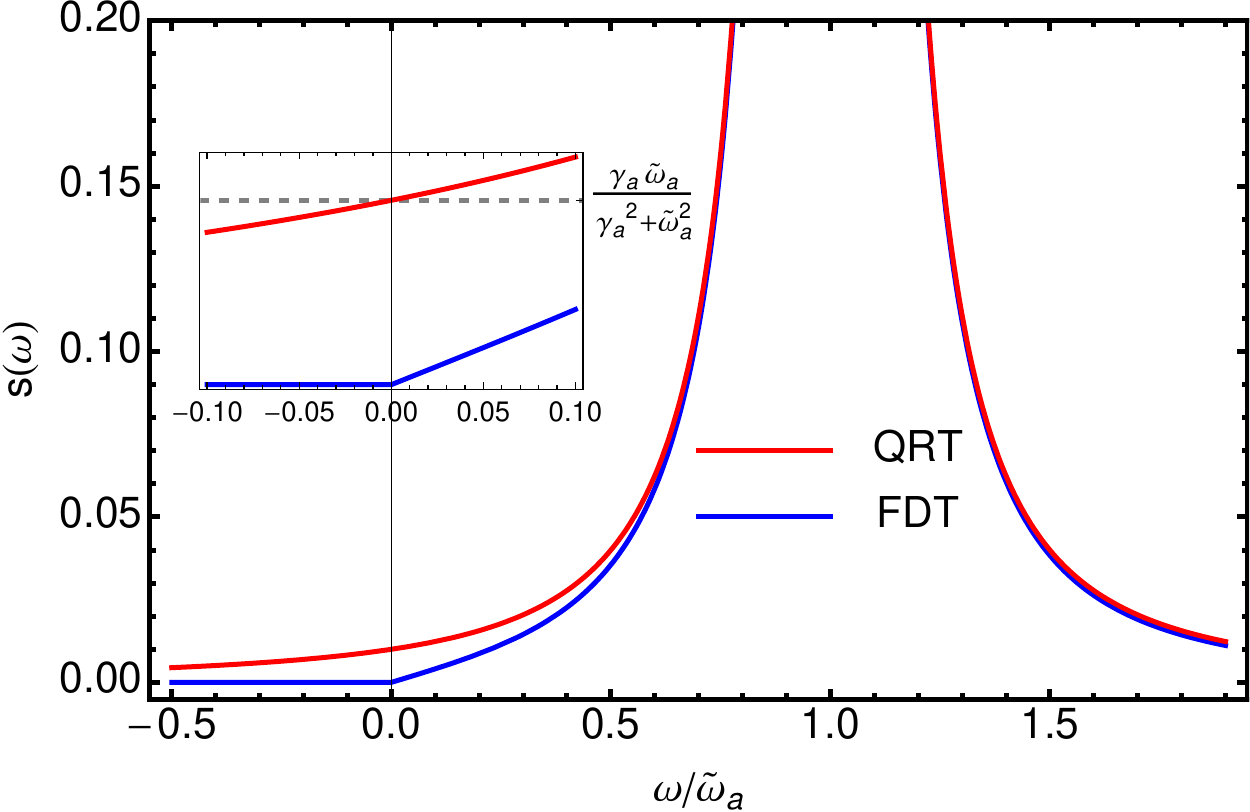}
\vspace{-.3cm}
\caption{Comparison of the normalized power spectra $s(\omega) = \underline{S} \cdot (\mathbf{d}\mathbf{d} / \pi \tilde{\omega}_{a})^{-1}$ 
predicted by the FDT (blue line) and the QRT (red line) for the polarizability of Eq. \eqref{alpha2ndOrder} (dissipation is set to
$\gamma_a/\tilde{\omega}_a=10^{-2}$).
While the two results are indistinguishable on resonance, they differ at low and negative frequencies. This is clearly visible in the inset, which provides a close up 
around $\omega=0$. While the power spectrum given by the FDT vanishes for $\omega \le 0$, that obtained from the QRT is nonzero. }
\label{PowerSpectrumComp}
\end{figure}

The above discussion shows that for large $\tau$ the equilibrium correlation function predicted by the FDT 
exhibits a power-law decay instead of an exponential behavior (as would result from the QRT  (see Eq.(\ref{QRT-TLS})). 
The power spectrum obtained from the QRT dipole correlation in  Eq. \eqref{QRT-TLS},
\begin{equation}
\underline{S}_{\rm QRT}(\omega) = \frac{\mathbf{d}\mathbf{d}}{\pi \tilde{\omega}_{a}}  \frac{\gamma_{a}\tilde{\omega}_{a}}{(\tilde{\omega}_{a}-\omega)^{2}+\gamma^{2}_{a}},
\end{equation}
is incorrect at low frequencies: It does not vanish for $\omega\le 0$ unlike the FDT power spectrum  (Eqs. \eqref{FDT_general} and \eqref{alpha2ndOrder}).
In Fig. \ref{PowerSpectrumComp} we compare $\underline{S}_{\rm QRT}(\omega)$ with the corresponding FDT power spectrum $\underline{S}(\omega)$. We see that the largest difference occurs in the region $\omega\sim 0$, while around the resonance value the two expressions overlap. Notice that the impact of this inaccurate description and the resulting discrepancies only appear for evaluations at orders higher than the second, i.e. when radiative damping induced by the interaction with electromagnetic field is nonzero.


In the next section, we will discuss the quantum frictional force experienced by an atom flying parallel to 
the surface. Since the evaluation of quantum friction requires a calculation at least to the fourth order 
in the atom-field coupling, the above discussion about the FDT vs the QRT will play a relevant role in 
determining the correct expression for this force.


\section{Non-Markovian effects in quantum friction} 
\label{dynamical}

In the previous section we have shown how non-Markovianity impacts the value of the static atom-surface  interaction. The relevance of the correction with respect to the Markovian case 
quantitatively depends on the level of accuracy used in describing the atom's internal dynamics, and it 
appears only when orders higher than the leading one are considered. In this section we move a step 
forward and generalize our analysis to an atom moving with constant velocity $\mathbf{v}$ parallel to a 
planar surface. Our goal is to evaluate the impact of non-Markovian effects on the vacuum-mediated 
frictional force experienced by the atom during its motion parallel to the surface. For simplicity we will 
consider the case $T=0$, i.e. quantum friction. 

\subsection{Derivation of the quantum frictional force}

\begin{figure}[t]
\includegraphics[width=8.5cm]{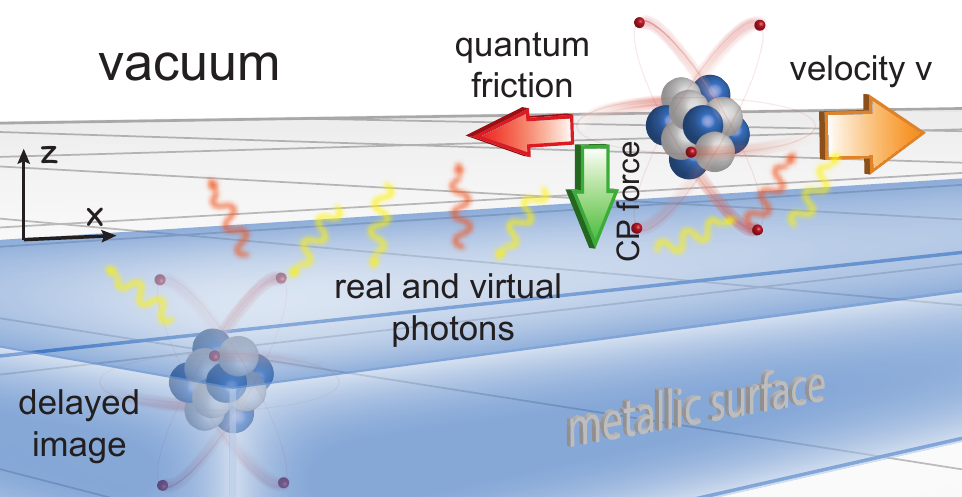}
\vspace{-.3cm}
\caption{Schematic of quantum friction on an atom moving at constant velocity parallel to a surface.
As in the static case, the presence of the correlation tensor in the expression of the frictional force 
can be qualitatively understood as stemming from the interaction of the atom with its delayed image within the material.}
\label{friction}
\end{figure}

In a semiclassical approximation, the trajectory for the center of mass of the atom will be prescribed 
as $\mathbf{r}_a(t)$, and its internal degrees of freedom will be modeled as an electric dipole. For 
simplicity, we neglect any magnetic contribution, which is a good approximation as long as the electric 
dipole is very near to the surface (near field region) \cite{Dedkov02a}. Assuming that the surface lies in the plane $z=0$  
and that the motion takes place at a constant distance $z_{a}$ from the surface (see Fig. \ref{friction}), 
the equation of motion for the atom's center of mass is 
$m \ddot{\mathbf{r}_a}(t)=\mathbf{F}_{\rm ext}+\mathbf{F}_{\rm fric}(t)$, 
where
\begin{equation}
\mathbf{F}_{\rm fric}(t)=\langle \hat{d}_{i}(t) \nabla_{\|}\hat{E}_{i}(\mathbf{r}_a(t),t)\rangle .
\end{equation}
Here, $\nabla_{\|}\equiv(\partial_{x},\partial_{y})$ is the gradient parallel to 
the surface. We assume the existence of an external classical force $\mathbf{F}_{\rm ext}$ on the atom 
that enforces a prescribed trajectory $\mathbf{r}_a(t)=({\bf R}_a(t), z_a)$. For times $t<t_i$, the atom 
is static at $\mathbf{r}_a=({\bf R}_a,z_a)$ and then undergoes an acceleration for some interval of time.
Eventually, it reaches a non-equilibrium steady state (NESS) given by $\mathbf{r}_a(t)=({\bf R}_a+ {\bf v} t,z_a)$ 
(for a discussion of the influence of the boost on a perturbative calculation of quantum friction, we refer to 
\cite{Barton10b,Hoye14,Intravaia15}).


Once again, the total field exhibits a component $\mathbfh{E}_{0}$ that is only related to the dissipative 
medium. In the dynamical case, however, a separation into positive and negative frequency parts is 
inappropriate because these parts get mixed by the Doppler shift (see below). Therefore, we write 
$\mathbfh{E}_{0}(\mathbf{r}(t),t)=\mathbfh{E}^{\oplus}_{0}(\mathbf{r}(t),t)+{\rm h.c.}$ 
\cite{Scheel09,Barton10b}, where
\begin{equation}
\mathbfh{E}_{0}^{\oplus}(\mathbf{r}(t),t)=
\int_{0}^{\infty}\frac{d\omega}{2\pi}\int \frac{d^{2}\mathbf{k}}{(2\pi)^{2}}\mathbfh{E}_{0}(\mathbf{k},z_{a};\omega)e^{\imath[\mathbf{k}\cdot\mathbf{R}(t)-\omega t]} ~.
\label{E0Motion}
\end{equation}
For $\omega>0$, the function $\mathbfh{E}_{0}(\mathbf{k},z_{a}; \omega)$ contains the same annihilation 
operators as in the static case. 
As in the previous section, by again choosing normal ordering and taking the average 
over the same initial state, the contribution of the field operator $\mathbfh{E}_{0}^{\oplus}$ to the 
quantum frictional force vanishes identically 
\cite{Scheel09,Barton10b}.
In other words, considerations similar to those made for the static force after Eq. \eqref{EReal} are 
also valid for the dynamic case, leading to the frictional force 
\begin{multline}
\mathbf{F}_{\rm fric}(t)=
\mathrm{Re}\bigg(\frac{2\imath}{\pi}\int_{0}^{\infty}d\omega\, \int_{0}^{t-t_{i}}d\tau e^{-\imath \omega \tau}\int\frac{d^{2}\mathbf{k}}{(2\pi)^{2}}  \imath \mathbf{k}\\
\times\mathrm{Tr}\left[\underline{C}(t,t-\tau)\cdot \underline{G}^{\sf T}_{\Im}(\mathbf{k},z_{a}, \omega)\right]e^{\imath \mathbf{k}\cdot[\mathbf{R}_a(t)-\mathbf{R}_a(t-\tau)]}\bigg) ~.
\label{friction1}
\end{multline} 
Here, we have defined
$\underline{G}_{\Im}(\mathbf{k},z,\omega)
 =
 [\underline{G}(\mathbf{k},z,\omega)-\underline{G}^{\dag}(\mathbf{k},z,\omega)]/(2 \imath)
 = \underline{G}^{\rm s}_{I}(\mathbf{k},z,\omega)
 -\imath\underline{G}^{\rm as}_{R}(\mathbf{k},z,\omega)$. 
From the properties of the Green tensor (see Appendix \ref{MaxwellG}), we can also deduce that the 
symmetric part of the Green tensor is even in $\mathbf{k}$, 
while the antisymmetric part is odd in $\mathbf{k}$.

Owing to the dissipative properties of the system, we expect that it has a finite memory time $\tau_c$,
so that the largest contributions in the $\tau$-integral in Eq. (\ref{friction1}) stem from times 
$\tau=t-t' \lesssim  \tau_c$. In particular, this means that in the limit of large times $t\to \infty$  
which we will consider below, we are allowed to replace $\mathbf{R}(t)$ by $\mathbf{R}_{a}+\mathbf{v} t$, 
and $\mathbf{R}(t-\tau)$ can be approximated by $\mathbf{R}_{a}+\mathbf{v}(t-\tau)$.
At large times, the atom reaches a steady state, and the frictional force balances 
the external force, resulting in an atom that moves with constant velocity $\mathbf{v}$ above the surface. 
One of the main differences with respect to the static case is that now the stationary correlation tensor 
depends on the velocity of the atom: 
$\underline{C}(\tau) =\lim_{t\to \infty}\langle \mathbfh{d}(t)\mathbfh{d}(t-\tau)\rangle 
 =
 \mathrm{tr}\left[\hat{d}_{i}(0)\hat{d}_{j}(-\tau)\hat{\rho}_{\rm NESS}\right] 
 \equiv 
 \underline{C}(\tau; \mathbf{v})$.
Similar to the static case, owing to the stationarity of the process, the correlation tensor depends 
only on the time difference $\tau$. It also implicitly depends on the final velocity ${\bf v}$ through 
the dynamics of the atom. In the NESS, the frictional force becomes constant 
$\mathbf{F}_{\rm fric}(t\to\infty)
 =
 \mathbf{F}_{\rm fric}$, 
and takes the form
\begin{multline}
\mathbf{F}_{\rm fric}=-
\mathrm{Re}\bigg(\frac{2}{\pi}\int_{0}^{\infty}d\omega\int\frac{d^{2}\mathbf{k}}{(2\pi)^{2}}\mathbf{k}  \int_{0}^{\infty}d\tau e^{-\imath (\omega- \mathbf{k}\cdot\mathbf{v}) \tau} \\
\times\mathrm{Tr}\left[\underline{C}(\tau;\mathbf{v})\cdot \underline{G}^{\sf T}_{\Im}(\mathbf{k},z_{a}, \omega)\right]\bigg).
\label{friction2}
\end{multline} 
Again in analogy with the static case, we define the power spectrum tensor
\begin{equation}
\underline{S}(\omega; \mathbf{v}) = \frac{1}{2\pi}\int_{-\infty}^{\infty}d\tau e^{\imath\omega\tau} \underline{C}(\tau;\mathbf{v}).
\end{equation}
Since  $\underline{C}^{\dag}(\tau;\mathbf{v})=\underline{C}(-\tau;\mathbf{v})$, the power spectrum 
is a Hermitean tensor $\underline{S}^{\dag}(\omega;\mathbf{v})=\underline{S}(\omega;\mathbf{v})$. This 
means that its real part is a symmetric tensor and its imaginary part is an anti-symmetric tensor. Using 
these symmetry properties of the power spectrum, together with those of $\underline{G}_{\Im}$, we can 
rewrite Eq.\eqref{friction2} as the sum of two contributions, 
$\mathbf{F}_{\rm fric}
 =
 \mathbf{F}^{\rm t}_{\rm fric}+\mathbf{F}^{\rm r}_{\rm fric}$, 
where
\begin{align}
\mathbf{F}^{\rm t}_{\rm fric}=&-
2\int_{0}^{\infty}d\omega\int\frac{d^{2}\mathbf{k}}{(2\pi)^{2}} \, \mathbf{k}\nonumber\\
&\times\mathrm{Tr}\left[\underline{S}_{R}(\mathbf{k}\cdot\mathbf{v}-\omega;\mathbf{v})\cdot \underline{G}^{\rm s}_{I}(\mathbf{k},z_{a}, \omega)\right],
\label{friction2t}
\end{align} 
and 
\begin{align}
\mathbf{F}^{\rm r}_{\rm fric}=
&
2\int_{0}^{\infty}d\omega\int\frac{d^{2}\mathbf{k}}{(2\pi)^{2}}  \, \mathbf{k}\nonumber\\
&\times\mathrm{Tr}\left[\underline{S}_{I}(\mathbf{k}\cdot\mathbf{v}-\omega;\mathbf{v})\cdot \underline{G}^{\rm as}_{R}(\mathbf{k},z_{a}, \omega)\right].
\label{friction2r}
\end{align}  
For simplicity, we model the dipole operator again as $\mathbfh{d}(t)=\mathbf{d}\hat{q}(t)$.
In this case $C_{ij}(\tau;\mathbf{v}) = d_{i}d_{j}\langle\hat{q}(\tau)\hat{q}(0)\rangle_{\rm NESS} = C_{ji}(\tau;\mathbf{v})$, 
i.e., the correlation tensor is symmetric. As a consequence, the power spectrum is also a symmetric 
tensor, and hence $\underline{S}_{I}(\omega;\mathbf{v})=0$. This implies that for this model $\mathbf{F}^{\rm r}_{\rm fric}=0$. 
The physical meaning of the term $\mathbf{F}^{\rm r}_{\rm fric}$ will be discussed in a more general 
context in a future work.

At this point, and within our assumptions, Eq.(\ref{friction2t}) is exact. It gives the quantum 
frictional force on an atom that asymptotically moves at constant velocity parallel to the surface. 
In order to further evaluate this force, we need to compute the non-equilibrium power spectrum 
tensor $\underline{S}(\omega;{\bf v})$. Unfortunately, the calculation of the dipole power spectrum 
in the NESS is a complex problem which often can be addressed only within perturbation theory. In 
the following, we describe two different approaches based again on the FDT and QRT, which 
yield markedly different predictions for the quantum friction in the low-velocity limit. 
This allows us to assess the impact of non-Markovianity on non-equilibrium fluctuation-induced 
interactions.


\subsection{FDT in quantum friction}
\label{FTDQF}

Strictly speaking, it is not valid to use of the FDT for the problem of a moving atom above a surface since the 
system is not in equilibrium. Nevertheless, earlier works 
\cite{Dedkov02a,Maghrebi12,Maghrebi13a,Hoye14}  
have (implicitly or explicitly) relied on the FDT to calculate quantum friction, assuming that both 
the atom and the surface are \emph{locally} in thermal equilibrium (LTE). Although the LTE approximation has been 
used in the literature for several non-equilibrium systems  (e.g., radiative heat transfer 
\cite{Polder71} 
or static atom-surface and Casimir forces out-of-thermal equilibrium 
\cite{Obrecht07}), 
its justification is still a matter of discussion. We will discuss the validity of the LTE approximation in quantum friction in a forthcoming publication
\cite{Intravaia16a}.
Interestingly, however, despite the lack of general results, in the case of quantum friction it 
is still possible to draw general conclusions in the low-velocity limit \cite{Intravaia14}. 

For symmetry reasons an expansion of the friction force for low velocities must contain only odd powers of $v$.
As $ \underline{G}^{\rm s}_{I}(\mathbf{k},z_{a},\omega)$ is even in $\mathbf{k}$, the frictional 
force in Eq.\eqref{friction2t} identically vanishes for $\mathbf{v}=0$, as it should. Also, as the 
power spectrum explicitly depends on the wave vector only through the Doppler-shifted frequency 
$\omega'=\omega-\mathbf{k}\cdot\mathbf{v}$, only two terms can contribute to the drag force in 
the small velocity limit: The first term is that in which we set the explicit ${\bf v}$-dependence 
to zero in the Doppler shift and retain the implicit velocity-dependence through the NESS density 
matrix, i.e., $\underline{S}_{R}(-\omega; \mathbf{v})$. The second term is that in which we set 
${\bf v}=0$ in the implicit velocity-dependence and retain the Doppler shift, i.e., 
$\underline{S}_{R}(-\omega'; {\bf v}=0)$. The first term does not contribute 
to the low-velocity drag force because the integral over ${\bf k}$ in Eq.\eqref{friction2t} identically 
vanishes as $\underline{G}^{\rm s}_{I}(\mathbf{k},z_{a}, \omega)$ is even in $\mathbf{k}$. 
The factor $\underline{S}_{R}(-\omega'; {\bf v}=0)$ in the second term is 
identical to the static power spectrum with negative Doppler-shifted  frequency. Because it effectively 
corresponds to an equilibrium situation (${\bf v}=0$ in $\rho_{\rm NESS}$), its contribution to 
Eq.\eqref{friction2t} can be computed using the corresponding FDT
\begin{equation}
\underline{S}_{R}(-\omega';{\bf v}=0)=\frac{\hbar}{\pi}\theta(-\omega')\underline{\alpha}_{I}(-\omega'),
\label{FDTDoppler}
\end{equation}
where $\underline{\alpha}$ is the polarizability tensor for the atom \emph{at rest}. Eq. \eqref{FDTDoppler} 
is a simplification of Eq. \eqref{FDT_general}, where we have removed the superscript "${\rm s}$" because 
in our dipole model the polarizability and the dipole correlator are symmetric by assumption. 

Since the surface is invariant under rotations around the $z$-axis, in the following we assume without 
loss of generality that the surface-parallel motion occurs in the $x$-direction, i.e., $\mathbf{v}=v_{x}\mathbf{x}$ 
($\mathbf{x}$ is the unit vector along the $x$-direction), which, considering again the parity properties 
of the Green tensor, implies that $\mathbf{F}_{\rm fric} = F_{\rm fric}\mathbf{x}$. Based on the arguments 
in the previous paragraph, the linear order in the velocity is included in
\begin{multline}
F_{\rm fric}\approx
-\frac{2\hbar}{\pi}
\int_{-\infty}^{\infty}\frac{dk_{y}}{2\pi}\int_{0}^{\infty}\frac{dk_{x}}{2\pi} k_{x} \int_{0}^{k_{x}v_{x}}d\omega \\
\times \mathrm{Tr}\left[\underline{\alpha}_{I}(k_{x}v_{x}-\omega;0)\cdot \underline{G}_{I}(\mathbf{k},z_{a},\omega)\right] ~.
\label{frictionFDT}
\end{multline}

At this point it is important to clarify the role played by the different terms in the above expression. 
The cut-off in the $\omega$-integral and the restriction to positive $k_{x}$ is due to the Heaviside function 
$\theta(-\omega')=\theta(k_{x}v_{x}-\omega)$ in Eq. \eqref{FDTDoppler}: this enforces a vanishing power spectrum for positive Doppler-shifted frequencies, 
revealing an important part of the underlying physics involved in the quantum friction process. Indeed, in the 
NESS as well as in the static case, only positive frequencies contribute to the quantum dynamics that results 
from the energy exchange among the different components of the system. When $v_{x}=0$, $\omega'=\omega>0$ and 
the Heaviside $\theta(-\omega')$ function identically vanishes, explaining why the frictional force is zero at zero velocity. 
However, due to the motion, we have that $\omega'=\omega-k_{x}v_{x}<0$ in the interval $0<\omega<k_{x}v_{x}$, 
which results in a nonzero contribution to the integral in Eq.\eqref{frictionFDT}. 
The so-called anomalous Doppler-effect \cite{Frolov86,Ginzburg96} occurs in this region, where part 
of the kinetic energy of the atom is converted into real excitations.
This mechanism is very much 
related to the physics of the Vavilov-Cherenkov effect 
\cite{Cerenkov37,Frank37,Maslovski11b,Maghrebi13a,Pieplow15}.
The remaining terms in Eq.\eqref{frictionFDT} are connected to the strength of the energy exchange and they 
essentially describe the density of states for the atomic system ($\underline{\alpha}_{I}$) and for the 
electromagnetic field emitted by the surface ($\underline{G}_{I}$). 

Because they are odd, the 
imaginary parts of both susceptibilities vanish at $\omega=0$, and since the Green tensor limits the values 
of wave vectors to $|\mathbf{k}|\lesssim z_{a}^{-1}$, an expansion at the lowest order in $v_{x}$ gives 
\begin{equation}
F_{\rm fric} \approx  -\frac{2\hbar v_{x}^{3}}{3(2\pi)^{3}}
\int_{-\infty}^{\infty}\hspace{-.2cm}dk_{y}\int_{0}^{\infty}\hspace{-.2cm}dk_{x}\, k_{x}^{4} \mathrm{Tr}\left[\underline{\alpha}'_{I}(0)\cdot \underline{G}'_{I}(\mathbf{k},z_{a}, 0)\right].
\label{lowvelocity}
\end{equation}
Here, we have assumed that the first derivative in $\omega$ (indicated by the prime in the above expression) 
of both tensors does not vanish (Ohmic behavior). 
Note that when either $\underline{\alpha}'_{I}(0)$ or $\underline{G}'_{I}$ are zero, higher (odd) orders in 
$v_{x}$ appear in the expansion. A detailed quantitative evaluation of Eq. \eqref{lowvelocity} requires the 
atom's low-frequency polarizability, which is known within the limits of perturbation theory
\cite{Milonni04,Buhmann04,Lach12,Jentschura15} (see also Section \ref{2ndOrderParagraph}). 

The above arguments demonstrate that, within our description of the atom, independently of the details of the 
internal dynamics (linear or non-linear), the lowest-order expansion in velocity of the zero-temperature stationary 
frictional force on an atom moving above a planar surface is at least cubic in $v_{x}$. This outcome is in 
agreement with some of the results available in the literature 
\cite{Pendry97,Volokitin07}, 
and it is formally equivalent to the application of the LTE approximation. There are, however, a few points that 
distinguish our approach from the use of that approximation. First, the derivation of Eq. \eqref{lowvelocity} 
presented above provides a plausibility argument for the application of the LTE+FDT approach to the quantum friction problem.
Second, the dependence on $v_{x}^{3}$ is a consequence of a general property of the polarizability, the crossing relation, and its behavior at low frequencies. This shows that, 
at least for the case of surface-parallel motion, the low velocity behavior of quantum friction is not 
related to the details of the (velocity-modified) internal dynamics, e.g. velocity dependent damping rates or level shifts \cite{Scheel09,Klatt16}. Third, in Eq. \eqref{lowvelocity}, 
there can be other $v_{x}^{3}$ contributions to the frictional force arising from the intrinsic non-equilibrium 
velocity-dependence of the power spectrum $S(\omega;{\bf v})$. If such contributions existed, they 
would imply a failure of the LTE approximation in the treatment of quantum friction. 
Indeed, we will show in a separate paper that such additional $v_{x}^{3}$-contributions do exist for the case of an atom modeled as a harmonic oscillator. 



\subsection{QRT in quantum friction}

We now compare to the friction force at low velocities predicted by the QRT formula. 
From the previous sub-section and \ref{FDTvsQRT}, we may expect a quite different behaviour because of the dependency of the polarizability at low frequencies.

Our starting point is Eq. \eqref{friction2}. Following our discussion in the previous sub-section, in the low-velocity limit one can approximate the correlator 
$\underline{C}(\tau;{\bf v})$ by the static correlator $\underline{C}(\tau;{\bf v}=0)$, and only retain the 
velocity-dependence in the Doppler shift (the $\exp[-\imath(\omega-{\bf k} \cdot {\bf v})\tau]$ factor in 
Eq.\eqref{friction2}). As a result, in order to compute the frictional force Eq.\eqref{friction2}, we need 
to evaluate the quantity
\begin{multline}
\mathrm{Re}\int_{0}^{\infty}d\tau e^{-\imath(\omega-\mathbf{k}\cdot\mathbf{v})\tau}\underline{C}(\tau,0)=\\
\underbrace{\hbar\mathrm{Re}\sum_{i}\left(\frac{\imath\mathrm{ Res[\underline{\alpha}(\Omega_{i})]}}{\Omega_{i}+\omega-\mathbf{k}\cdot\mathbf{v}}
\right)}_{\underline{S}_{\rm QRT}(\mathbf{k}\cdot\mathbf{v}-\omega)}
\\ \underbrace{-
\hbar\mathrm{Re}\sum_{i}\left(
\frac{\imath\mathrm{ Res[\underline{\alpha}(\Omega_{i})]}}{\Omega_{i}+|\omega-\mathbf{k}\cdot\mathbf{v}|}\right)}_{\underline{S}_{\rm nM}(\mathbf{k}\cdot\mathbf{v}-\omega)}.
\label{quantity}
\end{multline}
For simplicity, as before, we considered the case where both correlation 
and the polarizability tensors are symmetric, dropping all superscripts.
The static correlation tensor, computed with the FDT, is given in Eq.\eqref{exactC}. We recall that the 
first term of the sum in  Eq.\eqref{exactC} corresponds to the exponential decay and is the result one 
would have obtained via the QRT expression for the static correlator (such as Eq. \eqref{QRT-TLS} for the 
two-level system). After performing the relevant integral, this term gives rise to the contribution 
$\underline{S}_{\rm QRT}(\mathbf{k}\cdot\mathbf{v}-\omega)$ in Eq.\eqref{quantity}. The second contribution  
in Eq.\eqref{quantity} arises from the last term of the correlator in Eq.\eqref{exactC}, which contains 
the non-Markovian (nM) behavior and yields the long time deviation from the decaying exponential. Assuming 
that the polarizability is a symmetric tensor, we can formally write 
$\underline{\alpha}(\omega)= \sum_i \{ {\rm Res}[\underline{\alpha}(\Omega_i)]/(\omega-\Omega_i) - 
{\rm Res}[\underline{\alpha}(\Omega_i^*)]/(\omega+\Omega_i^*) \}$, from which we obtain
\begin{equation}
\frac{ \underline{\alpha}(\imath\xi)-\underline{\alpha}(-\imath\xi)}{2\imath} =
- \xi \sum_i \left\{
\frac{ {\rm Res}[\underline{\alpha}(\Omega_i)]}{\xi^2+\Omega_i^2} -
\frac{ {\rm Res}[\underline{\alpha}(\Omega_i)]^*}{\xi^2+\Omega_i^{*2}} 
\right\} ~.
\end{equation}
Performing the relevant integrals in Eq. \eqref{quantity}, we obtain $\underline{S}_{\rm nM}(\mathbf{k}\cdot\mathbf{v}-\omega)$.
This analysis reveals that the two terms $\underline{S}_{\rm QRT}$ and $\underline{S}_{\rm nM}$ are very 
similar, and cancel each other for $\omega'=-(\omega-\mathbf{k}\cdot\mathbf{v})<0$. We shall see now that 
both contributions are relevant for the final expression of the frictional force.

Let us consider the first term $\underline{S}_{\rm QRT}$ in Eq.\eqref{quantity} and derive the 
corresponding QRT form of quantum friction at low velocities. Using that $|\mathbf{k}|\lesssim z_{a}^{-1}$
and again assuming without loss of generality that the motion is along the $x$-axis, to the lowest order 
in $v_{x}$ ($v_{x}\ll \min[|\Omega_{i}|] z_{a}$), we obtain
\begin{multline}
F^{\rm QRT}_{\rm fric}\approx-\frac{4\hbar v_{x}}{(2\pi)^{2}}
\int_{-\infty}^{\infty}dk_{y}\int_{0}^{\infty}dk_{x}\, k_{x}^{2} \\
\times\mathrm{Tr}\left[\sum_{i}\mathrm{Re}\left[\frac{\mathrm{\imath Res[\underline{\alpha}(\Omega_{i})]}}{(\Omega_{i}+\omega)^{2}}\right]\cdot \underline{G}_{I}(\mathbf{k},z_{a}, \omega)\right].
\label{frictionQRT}
\end{multline}
This means that at low velocities the QRT predicts a frictional force linear in the velocity of the atom 
\cite{Scheel09,Barton10b}. 
This result contrasts with the cubic dependence in $v_{x}$ in Eq. \eqref{frictionFDT}.

The difference between these two outcomes is due to the behavior of $\underline{S}(\omega)$ at low frequencies 
(see Section \ref{FDTvsQRT} and in particular Fig. \ref{PowerSpectrumComp}). Mathematically, this can be understood by considering the contribution of 
$\underline{S}_{\rm nM}(\mathbf{k}\cdot\mathbf{v}-\omega)$ to Eq.\eqref{friction2}: A direct calculation 
of the frictional force that originates from this term, results in an expression that
leads to a linear-in-$v_x$ part which cancels the contribution in Eq.\eqref{frictionQRT} (see Appendix 
\ref{FricNM}). 
Thus, a higher-order expansion is required and, by symmetry, the next order to be considered is cubic in 
$v_{x}$. The non-Markovian correction in Eq.\eqref{quantity} equals the first term in the sum for 
$\omega'=\omega-\mathbf{k}\cdot\mathbf{v}>0$, which is a direct consequence of a vanishing  power 
spectrum for these frequencies (see also Eq.\eqref{FDTDoppler}). 
In direct connection with the Wigner-Weisskopf approximation, we have indeed that $\underline{S}_{\rm QRT}(\mathbf{k}\cdot\mathbf{v}-\omega)$ 
contains frequencies which are not allowed and these are responsible for the linear-in-$v_x$ behavior 
seen in Eq.\eqref{frictionQRT}. 
The main effect of $\underline{S}_{\rm nM}(\mathbf{k}\cdot\mathbf{v}-\omega)$ is to limit the range of 
integration over frequency to $k_{x}v_{x}$ as in Eq. \eqref{frictionFDT}. In addition, since 
$\underline{S}_{\rm QRT}(0)+\underline{S}_{\rm nM}(0)=0$ and, again, $\underline{G}_{I}(\mathbf{k},z_{a},0)=0$ 
any further possible linear-in-$v_x$ terms will vanish and an expansion for low velocities 
leads to Eq.\eqref{lowvelocity}.


\section{Quantum friction to second-order in perturbation theory}
\label{2ndOrderParagraph}

For a better understanding of the behavior of the frictional force, it is convenient to study the atom-surface frictional force computed within 
perturbation theory in the electric dipole moment $\mathbf{d}$. The most relevant point of this sub-section is that the perturbative order relevant 
for a correct evaluation of the frictional force depends on the presence or the absence of dissipation 
in the system. In particular, a special role is played by dissipation in the particle's internal dynamics, 
with differences occurring depending on whether this dissipation is \emph{intrinsic} or \emph{induced} (i.e. radiative) by 
the interaction with the electromagnetic field. In systems with intrinsic dissipation (e.g. nano-particles), 
an approach at second order in the field-dipole coupling provides the leading contribution. For induced 
dissipation, we will see below that, at second order, quantum friction is instead exponentially suppressed 
at low velocities. In this case, relevant for atoms, the leading contribution in the low-velocity limit arises 
from fourth order term. The calculation of this order will require extra care: 
as discussed at the end of Sec. \ref{static}, at fourth order in the perturbative expansion non-Markovian effects become important. 
Including them is crucial to obtain the correct result for quantum friction.

Within a perturbative framework at second order in the dipole moment $\mathbf{d}$, the dipole correlator 
$\underline{C}(\tau, \mathbf{v})$, and therefore the non-equilibrium 
power spectrum entering in Eq. \eqref{friction2t}, loses its explicit dependence on the velocity. At this 
order, the correlation and the linear response tensors are calculated starting from the free evolution of 
the dipole operator: The system is decoupled from the electromagnetic field and, therefore, is locally in 
thermal equilibrium. This means that the FDT can be employed, and the frictional force takes the form of 
Eq. \eqref{frictionFDT}, where the tensor $\underline{\alpha}$ must be replaced with the bare polarizability. 
Inserting
$\underline{\alpha}_{I}^{(0)}(\omega)
 =
 (\mathbf{d}\mathbf{d}/\hbar)\pi\left[\delta(\omega_{a}-\omega)-\delta(\omega_{a}+\omega)\right]$ 
in Eq.\eqref{frictionFDT} gives
\begin{multline}
F^{(2)}_{\rm fric}\approx
-2
\int_{-\infty}^{\infty}\frac{dk_{y}}{2\pi}\int_{\omega_{a}/v_{x}}^{\infty}\frac{dk_{x}}{2\pi} \,k_{x} \\
\times \mathrm{Tr}\left[\mathbf{d}\mathbf{d}\cdot \underline{G}_{I}(\mathbf{k},z_{a}, k_{x}v_{x}-\omega_{a})\right].
\label{friction2nd}
\end{multline}
This expression indicates that only wave vectors $k_{x}>\omega_{a}/v_{x}$ contribute to the frictional 
force. However, since the relevant part of Green tensor is proportional to $\exp[-2 k z_{a}]$ with $k=\abs{\mathbf{k}}$ (see Appendix \ref{MaxwellG}), $F^{(2)}_{\rm fric}$ exponentially vanishes in the low velocity limit. 
To see this more clearly, we recall that the total Green tensor can be decomposed  as 
$ \underline{G}= \underline{G}_{0}+ \underline{g}$. 
Because of Lorentz invariance, the  vacuum contribution $ \underline{G}_{0}$ does not 
contribute to the frictional force 
\cite{Dedkov03,Volokitin08}.
The scattered part of the Green tensor $\underline{g}$ has a symmetric part (the only relevant part in 
Eq.\eqref{friction2nd}) whose imaginary part in the near-field regime can be written as (see Appendix \ref{MaxwellG})
\begin{equation}
\underline{g}_{I}(\mathbf{k},z_{a};\omega) = \frac{r_{I}(\omega)}{2\epsilon_{0}}ke^{-2k z_{a}} \left(\frac{k_{x}^{2}}{k^{2}}\mathbf{x}\mathbf{x}+\frac{k_{y}^{2}}{k^{2}}\mathbf{y}\mathbf{y}+\mathbf{z}\mathbf{z}\right) ,
\label{nearfieldG}
\end{equation}
where $k=|\mathbf{k}|=\sqrt{k_{x}^{2}+k_{y}^{2}}$ and $\epsilon_{0}$ is the vacuum permittivity. Further, 
$r(\omega) = [\epsilon(\omega)-1]/[\epsilon(\omega)+1]$ is the quasi-static approximation of the transverse 
magnetic (TM) reflection coefficient for the planar surface -- here, we recall that in the near-field regime 
($z_{a}\ll c/\omega$) only the TM polarization matters for dielectric or metallic surfaces, and we can neglect the $k$ dependence 
of the reflection coefficient (For simplicity, throughout this manuscript we neglect spatial dispersion).

\begin{figure}[t]
\includegraphics[width=8.5cm]{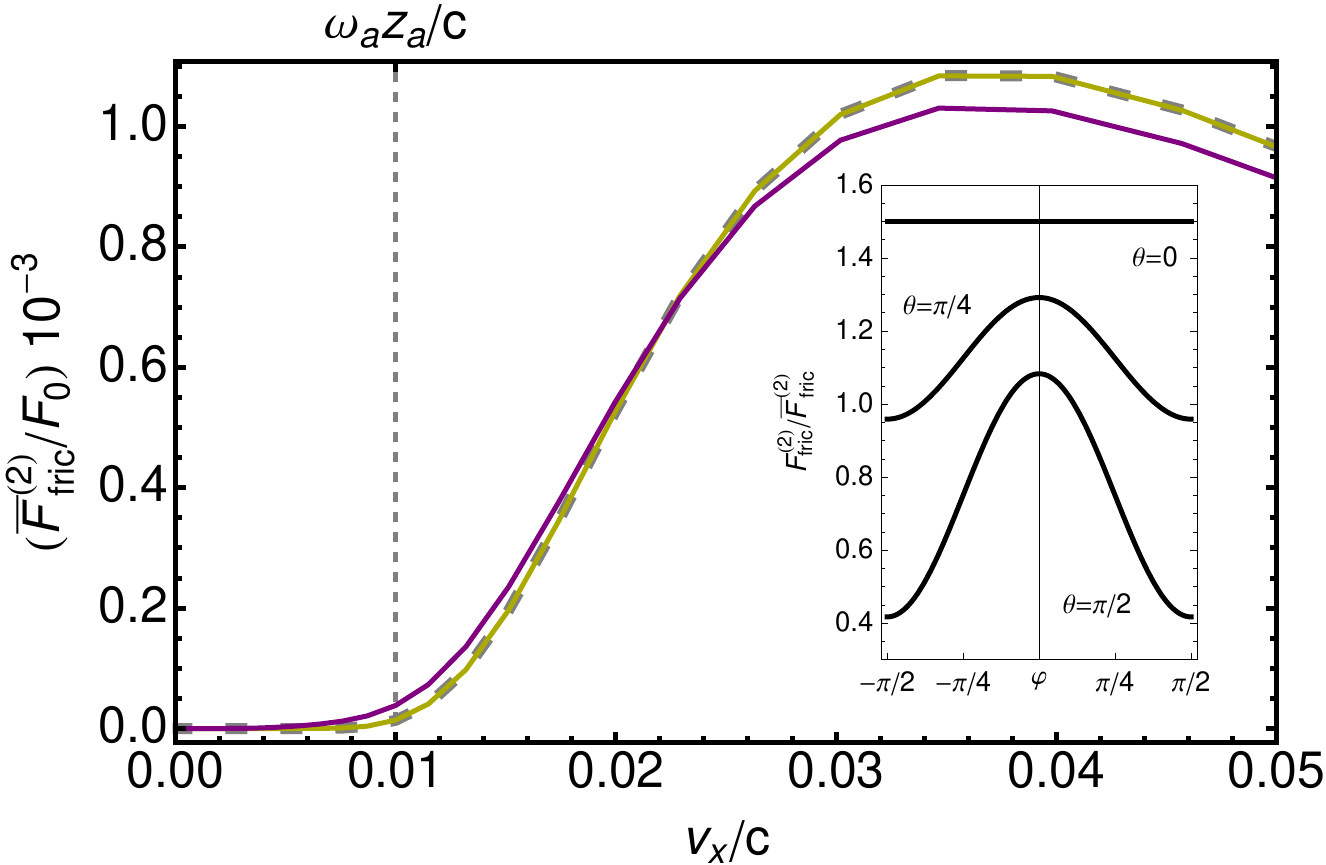}
\caption{
Velocity dependence of the normalized averaged second-order frictional force for a particle without intrinsic dissipation (e.g. an atom) moving 
above a metallic surface. The normalization is $F_{0}= -3\hbar \omega_{\rm sp}^5 \alpha_{0} / (2 \pi \epsilon_{0}c^4)$. The metal is described by the Drude model and results for various dissipation 
parameters are depicted: 
$\Gamma=0$ (black dashed, Eq. \eqref{friction2ndPlasma}), 
$\Gamma/\omega_{\rm sp}=10^{-3}$ (dark yellow), and
$\Gamma/\omega_{\rm sp}=10^{-1}$ (purple).
Further parameters are $\omega_{a}/\omega_{\rm sp}=0.2$ and $z_{a}\omega_{\rm sp}/c=0.05$, corresponding 
to the near-field regime, and $v_{x}/c=0.04$. 
The inset depicts the dependence of quantum friction for $\Gamma=0$, Eq. \eqref{friction2ndPlasma}, on the orientation 
of the vector of the dipole moment: The value $\theta=0$ corresponds to a dipole oriented normal to the surface; if tilted, the force
has is largest value is found for a dipole parallel to the plane defined by the motion and the normal to the surface ($\phi=0$). }
\label{friction2ndOr}
\end{figure}

Let us examine the case of a metallic surface described by the Drude model $\epsilon(\omega)=1-\omega_{p}^{2}[\omega(\omega+\imath \Gamma)]^{-1}$, where $\omega_p$ is the plasma frequency and $\Gamma$ the relaxation rate. In the limit of very small dissipation we have
$r_I(\omega) \approx (\pi\omega_{\rm sp}/2) \left[\delta(\omega-\omega_{\rm sp})-\delta(\omega+\omega_{\rm sp})\right]$,
where $\omega_{\rm sp}=\omega_{p}/\sqrt{2}$ is the surface plasmon resonance. In this low-dissipation limit, 
the integrals in Eq.\eqref{friction2nd} can be evaluated exactly, and the resulting second-order frictional 
force is 
\begin{equation}
\frac{F^{(2)}_{\rm fric}}{F_{0}} =
\frac{\omega_a/\omega_{\rm sp}}{12  \left( v_x/c \right)^{4}}  \left( 1+\frac{\omega_a}{\omega_{\rm sp}} \right)^3 
\mathcal{K}(u,\varphi,\theta)~, 
\label{friction2ndPlasma}
\end{equation}
where
\begin{equation}
\mathcal{K}(u,\varphi,\theta)=A_{0}(\varphi,\theta)K_{0}(2\abs{u})+A_{2}(\varphi,\theta)K_{2}(2\abs{u}) .
\label{callK}
\end{equation}
In these expressions, $u=(z_a \omega_{\rm sp}/v_x) ( 1+\omega_a/\omega_{\rm sp})$, $\theta$ and $\varphi$ are respectively the polar and azimuthal spherical angles describing the dipole vector ${\bf d}$.
In addition, $A_{0}(\varphi,\theta) =(3/2)\left[1+\left(3 \cos^2(\varphi)- 2\right)\sin^{2}(\theta)\right]$,
$A_{2}(\varphi,\theta) = (3/2)\left[1-\cos^{2}(\varphi)\sin^{2}(\theta)\right]$, and $K_{n}(x)$ is the 
modified Bessel function of the second kind and order $n$. The normalization is
$F_{0}= -3\hbar \omega_{\rm sp}^5 \alpha_{0} / (2 \pi \epsilon_{0}c^4)$, 
where $\alpha_{0}=2|\mathbf{d}|^{2}/(3\hbar \omega_{a})$ is the static isotropic atomic polarizability.
As an example, for using a plasma frequency $\omega_{p}= 9$ eV, and a $^{87}$Rb atom
($m=1.44 \times 10^{-25}$ kg, $\alpha_0=5.26 \times 10^{-39} {\rm F} {\rm m}^2$  \cite{Steck08}),  
we have $F_{0} \sim 0.31$ fN, corresponding to an acceleration  $F_0/m \sim 2.17 \times 10^{9}$ m/s$^2$.
Equation \eqref{friction2ndPlasma} 
demonstrates that the frictional force depends on the orientation of the dipole vector 
of the atom. The largest value of $F^{(2)}_{\rm fric}$ is found for dipoles 
oriented normal to the surface ($\theta=0$) or, if tilted with respect to the normal, when the vector of
the dipole moment exhibits a large component along the direction of the motion $\phi = 0$. 
The minimum value occurs for dipoles oriented along the $y$-axis, i.e., when the vector of the dipole moment 
is perpendicular  to both, the surface normal and the atom's propagation direction (see inset of Fig.\ref{friction2ndOr}). 
For simplicity, in the following we will consider expressions averaged over all dipole orientations, 
hence, $\bar{A_{0}}=\bar{A_{2}}=1$. 

Since $K_{n}\left(x\gg 1\right) \approx e^{-x}\sqrt{\pi/ 2x}$,  
the frictional force Eq.\eqref{friction2ndPlasma} is exponentially small at low velocities $v_x \ll z_a \omega_a$, namely
\begin{equation}
\frac{\bar{F}^{(2)}_{\rm fric}}{F_{0}}  \approx 
\sqrt{\frac{\pi  \left(\frac{\omega_a}{12\omega_{\rm sp}}\right)^{2}\left( 1+ \frac{\omega_a}{\omega_{\rm sp}} \right)^{5}}{\frac{\omega_{\rm sp}z_{a}}{c}\left(\frac{v_x}{c}\right)^{7}} } 
e^{-  \left(1+ \frac{\omega_a}{\omega_{\rm sp}} \right)  \frac{2 z_a \omega_{\rm sp}}{v_x}} .
\label{asympt1}
\end{equation}
In Fig. \ref{friction2ndOr} we display the velocity dependence of the second-order quantum frictional 
force for a substrate with vanishingly small dissipation, Eq.\eqref{friction2ndPlasma}. 
It is not exponentially suppressed only for velocities $v_x>z_a \omega_a$, which follows from the constraint on the wave vector discussed above. 
A maximum occurs for $v_{x}\approx (4/7)(\omega_{a}+\omega_{\rm sp})z_{a}$, i.e., 
roughly when the Doppler-shifted surface-plasmon resonance frequency is brought into resonance with 
the atomic transition (the reflection coefficient $r(v_{x}/z_{a}-\omega_{a})$ becomes large) \cite{Silveirinha14a}. This 
means that, at second order, quantum friction is essentially the result of a resonant process: The 
velocity must be sufficiently large (within the non-relativistic approach used here) so that the 
Doppler effect becomes anomalous, and the corresponding shifted frequencies are sufficiently large in order to include 
the (sharp) atomic transition and excite a plasmon. The photon (plasmon) in the near field is then 
scattered out of its ground state, while the atom is temporarily excited. The first photon is then followed 
by a second one resulting from the atomic decay, and both are finally absorbed by the surface 
\cite{Intravaia15}.

\begin{figure}[t]
\includegraphics[width=8.5cm]{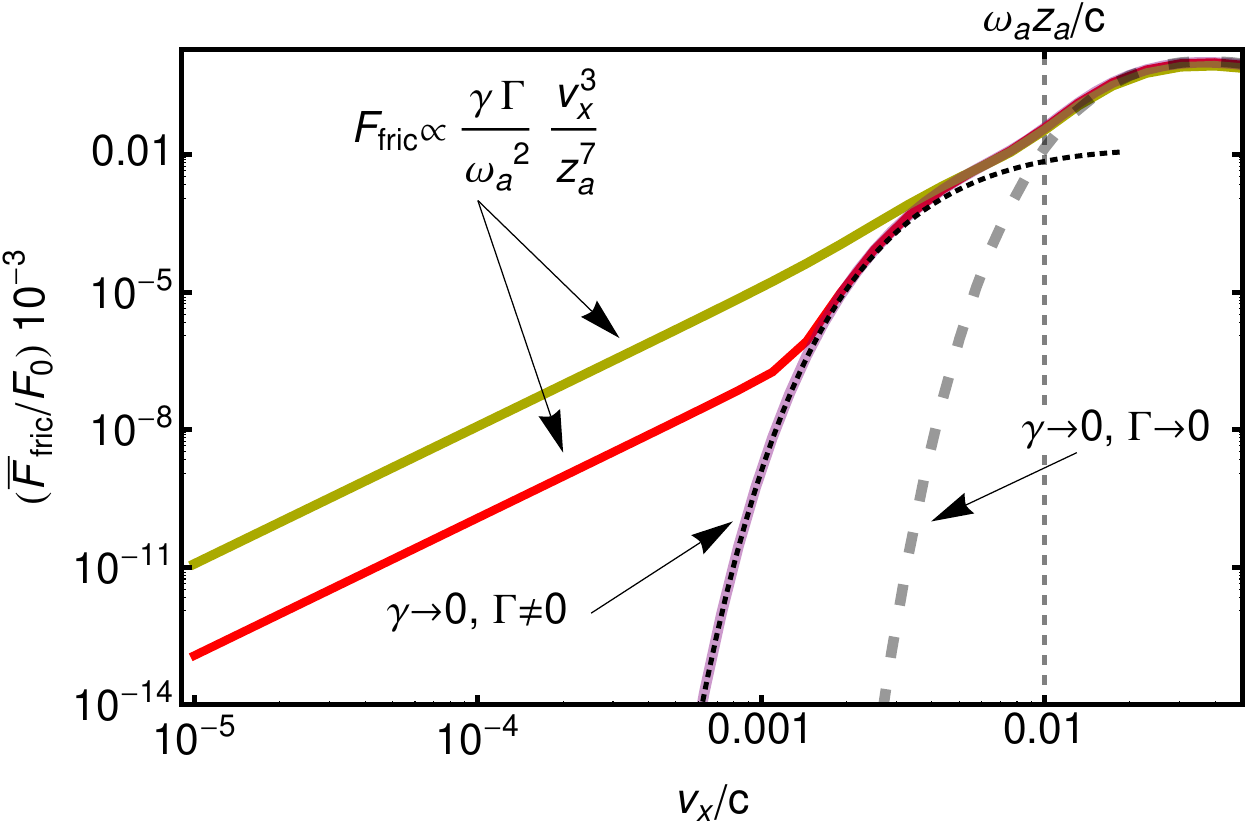}
\caption{
Velocity dependence of the normalized averaged second-order quantum frictional force for a particle with 
intrinsic dissipation. The normalization is $F_{0}= -3\hbar \omega_{\rm sp}^5 \alpha_{0} / (2 \pi \epsilon_{0}c^4)$. The particle has an internal resonance frequency $\omega_{a}/\omega_{\rm sp}=0.2$ 
and moves at a distance $z_{a}\omega_{\rm sp}/c=0.05$ above the surface, corresponding to the near-field 
regime. Results for two different values 
of the intrinsic dissipation are displayed: $\gamma/\omega_{\rm sp}=10^{-1}$ (dark yellow) and 
$\gamma/\omega_{\rm sp}=10^{-3}$ (red). In both cases, the damping $\Gamma$ is set to $\Gamma/\omega_{\rm sp}=10^{-1}$.
The purple curve shows the case for $\gamma=0$ and  $\Gamma/\omega_{\rm sp}=10^{-1}$. The black dotted 
curve represents the expression in Eq.\eqref{asympt2}. The thick gray dashed curve is Eq.\eqref{friction2ndPlasma}. 
}
\label{friction2ndOrDiss}
\end{figure}

When dissipation in the substrate is taken into account ($\Gamma \neq 0$), the second-order quantum 
frictional force essentially exhibits resonant behavior at high velocities ($v_x>z_a \omega_a$), as
discussed above (see Fig. \ref{friction2ndOr}).
At low velocities it still decays exponentially, but acquires a different asymptotic behavior due to 
the modification of the electromagnetic density of states, and is described by  
\begin{equation}
\frac{\bar{F}^{(2)}_{\rm fric}}{F_{0} }\approx \frac{\Gamma}{24 \omega_{\rm sp}}
\sqrt{\frac{ \left(\frac{\omega_a}{\omega_{\rm sp}}\right)^{7}}{\pi \left(\frac{\omega_{\rm sp} z_{a}}{c}\right)^{5}\left(\frac{v_{x}}{c}\right)^{3}}}
\left(1+\frac{5 v_x}{2 z_a \omega_a} \right) e^{-\frac{2 z_a \omega_a}{v_x}},
\label{asympt2}
\end{equation}
and is clearly visible in Fig. \ref{friction2ndOrDiss} (black dotted curve). 

Within the second-order perturbative approach, we can study the effect of dissipation associated with the 
particle's internal dynamics by considering the case of a system with intrinsic damping. Interestingly, 
the force on a moving object having intrinsic dissipation (e.g., a metallic nanoparticle) is qualitatively 
different from that on the moving atom. For this kind of system the bare polarizability is
$ \underline{\alpha}^{(0)}(\omega)=(2\mathbf{d}\mathbf{d}/\hbar\omega_{a}) \omega_{a}^{2}/(\omega_{a}^{2}-\omega^{2}-\imath\gamma\omega)$.
For the case of a metallic nanoparticle, $\omega_{a}=\omega^{\rm np}_{p}/\sqrt{3}$ is the resonance frequency 
of the localized surface plasmon, and $\omega^{\rm np}_{p}$ and $\gamma$ are, respectively, the plasma 
frequency and the dissipation rate for the bulk metal that comprises the particle.  
As demonstrated in Fig. \ref{friction2ndOrDiss}, when both sub-systems have finite dissipation, at low velocity
a further asymptotic appears in the force given in Eq.\eqref{frictionFDT}, and the frictional force is described by 
\begin{equation}
\frac{\bar{F}^{(2)}_{\rm fric}}{F_{0}} \approx   
\frac{45}{16} \frac{\Gamma}{24\pi}\frac{ \gamma }{\omega_a^2 } \left(\frac{c}{\omega_{\rm sp}z_a}\right)^7
\left(\frac{v_x}{c} \right)^3 ~.
\label{asympt3}
\end{equation}
Contrary to the previous cases, the second-order quantum frictional force for systems with intrinsic dissipation
\emph{does not} exponentially vanish 
at low velocities, but rather exhibits a cubic velocity dependence, just as predicted by Eq.\eqref{lowvelocity} 
on the basis of general arguments (both the polarizability and the Green tensor have a non vanishing first 
derivative with respect to $\omega$). 
In Fig. \ref{friction2ndOrDiss} we display all the three cases discussed above. Note that when both 
sources of dissipation are non-zero, the quantum frictional force approaches the results of the other 
cases for larger velocities. Specifically, the case $\Gamma\not=0$, $\gamma\to 0$ is approached first 
and this is followed by approaching the case $\Gamma,\,\gamma\to 0$. 

The three asymptotic expressions for low $v_{x}$ derived above cannot be obtained from each other by 
taking $\gamma \to 0$ or $\Gamma \to 0$, indicating that these limits do not commute and that, at low 
velocities, dissipation is very relevant for quantum frictional processes. This behavior is directly 
related to the corresponding increase in the density of states at low frequencies that is induced by 
dissipation (cf. the discussion following Eq. \eqref{frictionFDT}). Finally, although we have calculated 
the last two asymptotic expressions for a Drude metal, they can easily be generalized to other media 
having an Ohmic behavior at low frequencies, i.e., $r_{I}(\omega)\propto \omega$ for $\omega\to 0$ 
(in the computations above, we have used $r_{I}(\omega)\approx (\Gamma/\omega_{\rm sp}^{2})\, \omega$).


\section{Summary and Discussion} 

Using general concepts of quantum statistical mechanics, we have investigated the impact of 
non-Markovianity on  fluctuation-induced atom-surface interactions. 
In the static case, we have analyzed the failure of the Markov approximation by comparing the 
outcomes obtained using the FDT and the QRT. The FDT has led to a non-perturbative expression 
for the zero-temperature Casimir-Polder force, Eq.\eqref{imgFDT}. The FDT-based result contains the 
previously known expressions as special cases (see Appendix \ref{GenericCP} for details). 
We have shown that an alternative approach based on the QRT agrees with the FDT within lowest-order in perturbation theory in the atom-field coupling strength, but differ at higher orders. This can be intuitively understood by recalling 
that the QRT relies on the Born-Markov approximation, which works well for weak atom-field couplings 
and in a narrow range of frequencies close to resonance. This effectively limits the applicability 
of the QRT for fluctuation-induced interactions, given the broadband nature of the latter. 

The difference between the FDT and the QRT becomes even more pronounced when non-equilibrium 
effects are considered. Specifically, the distinct behavior at large times/low frequencies of the dipole-dipole correlation function 
is responsible for the different scaling laws of quantum friction with regard to atom velocity 
and atom-surface separation. 
Assuming for low frequencies an Ohmic behavior of the material. we have shown that, when (induced 
or intrinsic) dissipation is incorporated both, for the constituent material and the atom 
dynamics, quantum friction scales as $\propto v_x^{3}$ for low velocities. By contrast, a 
Markovian approach leads to a behavior $\propto v_x$. More generally, the exponent of the power 
law at low velocities is strongly related to the low frequency behavior of the atomic and the 
material susceptibilities. This explains why dissipation is so relevant: quantum friction is dominated by low frequencies
and there damping increases the density of states, thus opening new interaction 
channels.   

For systems that exhibit intrinsic dissipation (e.g. a nano-particle above a surface) a 
second-order perturbative calculation is sufficient to compute the leading term of low-velocity 
quantum friction (see Sec. \ref{2ndOrderParagraph}). However, for systems exhibiting only radiative damping (e.g., atoms), a 
consistent higher-order (fourth at least) calculation is required 
in order to accurately describe the quantum frictional process \cite{Barton10b,Intravaia15} . 

The approach presented in this manuscript is based on stationary systems and is thus not suitable 
for the analysis of non-steady-state configurations, e.g., an excited atom. However, it has 
recently been shown that in the case of non-stationary quantum friction the salient features of our 
results can also be recovered within a time-dependent perturbative approach to the fourth order in 
the atom-field coupling strength 
\cite{Intravaia15}. 
Intuitively, one can understand this result in terms of the characteristic time $\tau_{\rm NESS}$ 
which the system needs to reach the non-equilibrium steady state -- as even within a time-dependent 
perturbative scheme, steady-state features become relevant as soon as we consider $t\gg \tau_{\rm NESS}$.


\section{Acknowledgments}
We thank Gabriel Barton, Stefan Buhmann, Manuel Donaire, Belen Farias, Juliane Klatt, 
Kimball Milton, Stefan Scheel, and Wang-Kong Tse for useful discussions. We acknowledge 
support by the LANL LDRD program, and by the Deutsche Forschungsgemeinschaft (DFG) through 
project B10 within the Collaborative Research Center (CRC) 951 Hybrid Inorganic/Organic Systems 
for Opto-Electronics (HIOS).
FI further acknowledges financial support from the European Union Marie Curie People program 
through the Career Integration Grant No. PCIG14- GA-2013-631571. CH and FI acknowledge support from the DFG through the 
DIP program (grant FO 703/2-1 and SCHM 1049/7-1)


\appendix


\section{Properties of the Green tensor}
\label{MaxwellG}

The electromagnetic Green tensor 
is the solution of 
\begin{equation}
\nabla\times \nabla \times \underline{G}(\mathbf{r},\mathbf{r}'; \omega)-\frac{\omega^{2}}{c^{2}}\epsilon(\mathbf{r},\omega) \underline{G}(\mathbf{r},\mathbf{r}'; \omega)=\frac{\omega^{2}}{\epsilon_{0} c^{2}}\delta(\mathbf{r}-\mathbf{r}') ,
\label{GreenT}
\end{equation}
subject to appropriate boundary conditions. 
Some useful properties of the Green tensor are 
$\underline{G}^{\sf T}(\mathbf{r}_{1},\mathbf{r}_{2}; \omega) = \underline{G}(\mathbf{r}_{2},\mathbf{r}_{1}; \omega)$ (the 
superscript ${\sf T}$ indicates the transposed matrix) and
$\underline{G}(\mathbf{r},\mathbf{r}', \omega) = \underline{G}^{*}(\mathbf{r},\mathbf{r}',-\omega)$ 
which are, respectively, consequences of reciprocity and the fact that in the time domain its elements 
are real. Further, the Green tensor represents a susceptibility and, therefore, it is analytic in 
the upper half of the complex frequency plane and its real and imaginary parts satisfy Kramer-Kronig 
relations \cite{Jackson75}. As a first consequence, we have that ($\tau>0$)
\begin{multline}
\int_{-\infty}^{\infty}\frac{d\omega}{2\pi} \underline{G}(\mathbf{r}_{1},\mathbf{r}_{2}; \omega)e^{-\imath \omega\tau} =\\
\int_{-\infty}^{\infty}\frac{d\omega}{2\pi} \,\left[\mathrm{P}\int_{-\infty}^{\infty}\frac{d\nu}{\pi}\frac{\underline{G}_{I}(\mathbf{r}_{1},\mathbf{r}_{2}; \nu)}{\nu-\omega}+\imath \underline{G}_{I}(\mathbf{r}_{1},\mathbf{r}_{2}; \omega)\right]e^{-\imath \omega\tau}
\\
= 2\imath\int_{-\infty}^{\infty}\frac{d\omega}{2\pi}\, \underline{G}_{I}(\mathbf{r}_{1},\mathbf{r}_{2}; \omega)e^{-\imath\omega \tau}. 
\end{multline}
In the first line of the above calculation, we have used the Kramers-Kronig relations for the real part 
of the Green tensor, and in the second line we have employed the identity
\begin{equation}
\mathrm{P}\int_{-\infty}^{\infty}\frac{d\omega}{\pi}\frac{e^{-\imath\omega t}}{\nu-\omega}=\imath e^{-\imath\nu t} .
\end{equation}
In addition, in the main text we often encounter the (partial) spatial Fourier transform 
\begin{multline}
\underline{G}(\mathbf{r}_{1},\mathbf{r}_{2}; \tau)=\int_{-\infty}^{\infty}\frac{d\omega}{2\pi}\int \frac{d^{2}\mathbf{k}}{(2\pi)^{2}}\\
\times \underline{G}(\mathbf{k}, z; \omega)e^{\imath[\mathbf{k}\cdot(\mathbf{R}_{1}-\mathbf{R}_{2})-\omega\tau]} ,
\end{multline}
where $\mathbf{k}=(k_{x},k_{y})$ denotes the in-plane wave vector and the position vectors 
$\mathbf{r}_{1} = (\mathbf{R}_{1},z)$ and $\mathbf{r}_{2}=(\mathbf{R}_{2},z)$ feature the same 
$z$-coordinate. From the reality and reciprocity of the Green tensor, 
one gets
$\underline{G}^{\sf T}(\mathbf{k},z; \omega)=\underline{G}(-\mathbf{k},z; \omega)$, $\underline{G}^{*}(\mathbf{k},z; \omega)=\underline{G}(-\mathbf{k},z; -\omega)$, and therefore $\underline{G}^{\dag}(\mathbf{k},z; \omega)=\underline{G}(\mathbf{k},z; -\omega)$.
Based on these properties one can also deduce that the symmetric part of the Green tensor 
$\underline{G}^{\rm s}(\mathbf{k},z, \omega)$ is even in $\mathbf{k}$, while the antisymmetric part 
$\underline{G}^{\rm as}(\mathbf{k},z, \omega)$ is odd in $\mathbf{k}$. We can then write
\begin{align}
\underline{G}_{I}(\mathbf{r},\mathbf{r}',\omega)
= \int\frac{d^{2}\mathbf{k}}{(2\pi)^{2}} \,\underline{G}_{\Im}(\mathbf{k},z, \omega)e^{\imath \mathbf{k}\cdot(\mathbf{R}-\mathbf{R}')} ,
\end{align}
where we have defined the tensor
\begin{align}
\underline{G}_{\Im}(\mathbf{k},z,\omega)&=\frac{
\underline{G}(\mathbf{k},z,\omega)-\underline{G}^{\dag}(\mathbf{k},z,\omega)}{2 \imath}\nonumber\\
&= \underline{G}^{\rm s}_{I}(\mathbf{k},z,\omega)
-\imath\underline{G}^{\rm as}_{R}(\mathbf{k},z,\omega).
\end{align}

As explained in the main text, the total Green tensor can be decomposed  as 
$ \underline{G}= \underline{G}_{0}+ \underline{g}$, where $\underline{G}_{0}$ is the  vacuum contribution and
$\underline{g}$ is the scattered part.
For a planar surface, we have ($z>0$)
\begin{equation}
\underline{g}(\mathbf{k},z, \omega)=
\frac{\kappa}{2\epsilon_{0}}
\left(r^{p}[\omega,k]\mathbf{p}_{+}\mathbf{p}_{-}+ \frac{\omega^{2}}{c^{2}\kappa^{2}}r^{s}[\omega,k]\mathbf{s}\mathbf{s}\right)
e^{-2\kappa z} ,
\end{equation}
where $\kappa=\sqrt{k^{2}-\omega^{2}/c^{2}}$ ($k=|\mathbf{k}|$, $\mathrm{Re}[\kappa]>0$ and $\mathrm{Im}[\kappa]<0$), $\epsilon_{0}$ is the vacuum 
permittivity, and $r^{\sigma}[\omega,k]$ are the polarization dependent ($\sigma=s,p$) reflection 
coefficients of the surface. Furthermore, in the above expressions, we have defined the polarization 
vectors 
\cite{Wylie84} 
\begin{equation}
\mathbf{s}=\frac{\mathbf{k}}{k}\times \mathbf{z}\quad\mathbf{p}_{\pm}=\frac{k}{\kappa}\mathbf{z}\mp\imath\frac{\mathbf{k}}{k}~.
\end{equation}
The corresponding dyadic tensors can be then written as
 \begin{align}
\mathbf{s}\mathbf{s}=
\begin{pmatrix}
\frac{k_{y}^{2}}{k}&-\frac{k_{y}k_{x}}{k}&0\\
-\frac{k_{y}k_{x}}{k}&\frac{k_{x}^{2}}{k}&0\\
0&0&0
\end{pmatrix} ,
\end{align}
\begin{align}
\mathbf{p}_{+}\mathbf{p}_{-}=
\begin{pmatrix}
\frac{k_{x}^{2}}{k^{2}}&\frac{k_{y}k_{x}}{k^{2}}&-\imath \frac{k}{\kappa}\frac{k_{x}}{k}\\
\frac{k_{y}k_{x}}{k^{2}}&\frac{k_{y}^{2}}{k^{2}}&-\imath \frac{k}{\kappa}\frac{k_{y}}{k}\\
\imath \frac{k}{\kappa}\frac{k_{x}}{k}&\imath \frac{k}{\kappa}\frac{k_{y}}{k}&\frac{k^{2}}{\kappa^{2}}
\end{pmatrix} .
\end{align}

\section{Comparing various expressions for the Casimir-Polder force}
\label{GenericCP}

Following the procedure described in Section \ref{StaticCPFDT}, by employing the fluctuation dissipation 
theorem it is straightforward to show that the static Casimir-Polder can be written as
\begin{equation}
 F_{\rm CP}
=\frac{\hbar}{\pi}\mathrm{Im}
\int_{0}^{\infty}d\omega \, \mathrm{Tr}\left[\underline{\alpha}(\omega,\mathbf{r}_{a})\cdot \partial_{z}\underline{G}(\mathbf{r}_{a},\mathbf{r}, \omega)_{\vert \mathbf{r}=\mathbf{r}_{a}}\right] .
\label{finalCP}
\end{equation}
Despite formal similarities, Eq. \eqref{finalCP} differs from the standard formulas 
found in the literature for the atom-surface interaction. Our derivation of Eq. \eqref{finalCP} 
is non-perturbative \cite{Buhmann04} as it does not rely on any perturbative weak-coupling expansion. It is, therefore, 
different from the expression for the Casimir-Polder force obtained in second-order perturbation theory  
\cite{Casimir48a,Wylie84,McLachlan63}. 
In the latter, the position-independent polarizability $\underline{\alpha}^{(0)}(\omega)$ replaces 
the dressed, position-dependent polarizability $\underline{\alpha}(\omega,\mathbf{r}_{a})$ and thus
it corresponds to the lowest order of the perturbative expansion of Eq. \eqref{finalCP} in the coupling 
strength. Indeed, the bare polarizability represents the Fourier transform of the response function  
$\underline{\alpha}^{(0)}(\tau)=(\imath/\hbar)\theta(\tau)\mathrm{tr}[[\mathbfh{d}^{(0)}(\tau),\mathbfh{d}^{(0)}(0)]\hat{\rho}_a]$, 
where only the \emph{free} dipole evolution is considered ($\hat{\rho}_a$ describes the atomic 
ground state.) 

Perhaps more relevant is the difference between Eq. \eqref{finalCP} and the corresponding Lifshitz 
formula in the scattering formulation 
\cite{Dzyaloshinskii61,Lambrecht06,Intravaia11,Intravaia12b}, 
\begin{equation}
F^{\rm Lif}_{\rm CP}
=\frac{\hbar}{\pi} \;  \mathrm{Im}\int_{0}^{\infty} d\omega
\mathrm{Tr} \left[
\frac{ \underline{\alpha}_{\rm vac}(\omega) \cdot \partial_z  \underline{g}(\mathbf{r}_{a},\mathbf{r}, \omega)}
{1-\underline{\alpha}_{\rm vac} (\omega) \cdot  \underline{g}(\mathbf{r}_{a},\mathbf{r}, \omega)}
\right]_{\vert \mathbf{r}=\mathbf{r}_{a}}. 
\label{scattering}
\end{equation} 
In this case, the force is given in terms of the scattering properties of the surface and the atom 
is treated as a non-interacting scatterer: The quantity $\underline{\alpha}_{\rm vac}(\omega)$ is the 
position-independent dressed polarizability for the atom placed in the free electromagnetic vacuum 
(it differs from  $\underline{\alpha}^{(0)}(\omega)$ as it contains the Lamb shift and spontaneous 
decay rate), while $\underline{g}(\mathbf{r},\mathbf{r}', \omega)$ is again the scattered part of the Green 
tensor 
\cite{Intravaia11}. 

The denominator in the above expression indicates multiple reflections of the electric field between 
the atom and the surface. Keeping only one round-trip in the atom-surface multiple reflections process 
\cite{Messina09}, 
one obtains an expression which is formally similar to Eq.\eqref{finalCP} but where 
$\underline{\alpha}_{\rm vac}(\omega)$ replaces $\underline{\alpha}(\omega,\mathbf{r}_{a})$. When all 
multiple reflections are kept, Eq. (\ref{scattering}) becomes identical to Eq.\eqref{finalCP} provided 
that
\begin{equation}
\underline{\alpha}(\mathbf{r}_{a},\omega)=\left[1-\underline{\alpha}_{\rm vac}(\omega)\cdot\underline{g}(\mathbf{r}_{a},\mathbf{r}_{a},\omega)\right]^{-1}\cdot\alpha_{\rm vac}(\omega).
\label{relation}
\end{equation}
This corresponds to a particular re-summation of the perturbation series, and the underlying condition 
is that the atom responds linearly to the electric field. For example, when the atom is modeled as a 
linear harmonic oscillator system, the equation of motion of the dipole operator $\hat{\bf d} = {\bf d} \hat{q}$ 
is  
\begin{equation}
\ddot{\hat{q}}(t)+\omega_{a}^{2}\hat{q}(t)=\frac{2\omega_{a}}{\hbar}\mathbf{d}\cdot \mathbfh{E}(\mathbf{r}_{a},t)~.
\end{equation}
Here, $\mathbfh{E}(\mathbf{r},t)=\mathbfh{E}_{0}(\mathbf{r},t)+\mathbfh{E}_{\rm S}(\mathbf{r},t)$ is 
the sum of the field without the atom ($\mathbf{E}_{0}(\mathbf{r},t)$) and the field generated by the 
dipole ($\mathbf{E}_{\rm S}(\mathbf{r},\omega)=\underline{G}(\mathbf{r},\mathbf{r}_{a}, \omega)\cdot\mathbf{d}\hat{q}(\omega)$). 
In Fourier space, the stationary dipole's dynamics is then given by 
$\mathbfh{d}(\omega,\mathbf{r}_{a}) = \underline{\alpha}(\omega,\mathbf{r}_{a})\cdot\mathbfh{E}_{0}(\mathbf{r}_{a}, \omega)$ 
which implicitly defines the atom's polarizability as
\begin{equation}
\underline{\alpha}(\mathbf{r}_{a},\omega)=\frac{\frac{2\omega_{a}}{\hbar}\mathbf{d}\mathbf{d}}{\omega_{a}^{2}-\omega^{2}-\frac{2\omega_{a}}{\hbar}\mathbf{d}\cdot\underline{G}(\mathbf{r}_{a},\mathbf{r}_{a},\omega)\cdot\mathbf{d}} .
\label{genpol}
\end{equation}
This is the expression for the polarizability that enters in Eq.(\ref{finalCP}) which, as expected, 
depends on the position of the oscillator through the radiation reaction field represented by the 
Green tensor. If the atom is isolated in vacuum, one must replace $\underline{G}$  by $\underline{G}_{0}$ 
to recover $\alpha_{\rm vac}(\omega)$. Finally, using that  $\underline{G} = \underline{G}_{0}+\underline{g}$ 
one can show that, in the case of an oscillator, $\underline{\alpha}(\mathbf{r}_{a},\omega)$ and $\alpha_{\rm vac}(\omega)$ 
are related as described in Eq. \eqref{relation}.

\section{Using the QRT for calculating the Casimir-Polder force}
\label{ComplexFreq}

Within the QRT approach, the Casimir-Polder force is given as an integration over real frequencies 
via equation \eqref{sourceB}. It can be decomposed into four terms
\begin{subequations}
\begin{align}
F^{\rm QRT}_{\rm CP}
&=\frac{1}{2\pi\imath}\int_{0}^{\infty}d\omega\,  \frac{\mathrm{Tr}\left[\mathbf{d}\mathbf{d}\cdot \partial_{z}\underline{G}(\mathbf{r}_{a},\mathbf{r}, \omega)_{\vert \mathbf{r}=\mathbf{r}_{a}}\right]}{\omega+\tilde{\omega}_{a}(\mathbf{r}_{a})-\imath\gamma(\mathbf{r}_{a})}\label{a}\\
&-\frac{1}{2\pi\imath}\int_{0}^{\infty}d\omega\,  \frac{\mathrm{Tr}\left[\mathbf{d}\mathbf{d}\cdot \partial_{z}\underline{G}^{*}(\mathbf{r}_{a},\mathbf{r}, \omega)_{\vert \mathbf{r}=\mathbf{r}_{a}}\right]}{\omega+\tilde{\omega}_{a}(\mathbf{r}_{a})-\imath\gamma(\mathbf{r}_{a})} \label{b}\\
&+\frac{1}{2\pi\imath}\int_{0}^{\infty}d\omega\,  \frac{\mathrm{Tr}\left[\mathbf{d}\mathbf{d}\cdot \partial_{z}\underline{G}(\mathbf{r}_{a},\mathbf{r}, \omega)_{\vert \mathbf{r}=\mathbf{r}_{a}}\right]}{\omega+\tilde{\omega}_{a}(\mathbf{r}_{a})+\imath\gamma(\mathbf{r}_{a})} \label{c}\\
&-\frac{1}{2\pi\imath}\int_{0}^{\infty}d\omega\,  \frac{\mathrm{Tr}\left[\mathbf{d}\mathbf{d}\cdot \partial_{z}\underline{G}^{*}(\mathbf{r}_{a},\mathbf{r}, \omega)_{\vert \mathbf{r}=\mathbf{r}_{a}}\right]}{\omega+\tilde{\omega}_{a}(\mathbf{r}_{a})+\imath\gamma(\mathbf{r}_{a})}\label{d} ,
\end{align}
\end{subequations}
where $\tilde{\omega}_{a}(\mathbf{r}_{a})$ and $\gamma(\mathbf{r}_{a})$ are non-negative quantities.
The integrand in \eqref{c} has poles in the lower part of the complex-frequency plane. Thus, a standard Wick 
rotation can be performed, resulting in an integration over the positive imaginary frequency axis. For the 
integral in \eqref{a}, we notice that the integrand has all poles in the lower part of the complex-frequency 
plane with the exception of $\omega=-\tilde{\omega}_{a}(\mathbf{r}_{a})+\imath\gamma(\mathbf{r}_{a})$ which
is actually located in the upper left quadrant. This means that a similar Wick rotation in the first integrand 
is still possible.
For the term appearing in \eqref{b} we can utilize that 
$\underline{G}^{*}(\mathbf{r}_{a},\mathbf{r},\omega) = \underline{G}(\mathbf{r}_{a},\mathbf{r}, -\omega)$ 
and a change of variable $\omega\to -\omega$. The resulting integral running from $-\infty$ to zero 
concerns an integrand with poles in the lower part of the complex-frequency plane.  A similar procedure 
for the term in \eqref{d} leads to an integral where all the poles of the Green tensor are in the lower 
part of the complex-frequency plane, except for one pole at 
$\omega=\tilde{\omega}_{a}(\mathbf{r}_{a})+\imath\gamma(\mathbf{r}_{a})$ that is located in the first quadrant. 
Since the corresponding integral again runs from $-\infty$ to zero we can still perform a rotation of the 
complex path in the second quadrant. This procedure leads to
\begin{subequations}
\begin{align}
F^{\rm QRT}_{\rm CP}
&=\frac{1}{2\pi}\int_{0}^{\infty}d\xi\,  \frac{\mathrm{Tr}\left[\mathbf{d}\mathbf{d}\cdot \partial_{z}\underline{G}(\mathbf{r}_{a},\mathbf{r}, \imath\xi)_{\vert \mathbf{r}=\mathbf{r}_{a}}\right]}{\tilde{\omega}_{a}(\mathbf{r}_{a})-(-\imath\xi+\imath\gamma(\mathbf{r}_{a}))} \nonumber \\
&+\frac{1}{2\pi}\int_{0}^{\infty}d\xi\,  \frac{\mathrm{Tr}\left[\mathbf{d}\mathbf{d}\cdot \partial_{z}\underline{G}(\mathbf{r}_{a},\mathbf{r}, \imath\xi)_{\vert \mathbf{r}=\mathbf{r}_{a}}\right]}{\tilde{\omega}_{a}(\mathbf{r}_{a})-(\imath\xi+\imath\gamma(\mathbf{r}_{a}))} 
\nonumber \\
&+\frac{1}{2\pi}\int_{0}^{\infty}d\xi\,  \frac{\mathrm{Tr}\left[\mathbf{d}\mathbf{d}\cdot \partial_{z}\underline{G}(\mathbf{r}_{a},\mathbf{r}, \imath\xi)_{\vert \mathbf{r}=\mathbf{r}_{a}}\right]}{\tilde{\omega}_{a}(\mathbf{r}_{a})+(\imath\xi+\imath\gamma(\mathbf{r}_{a}))} 
\nonumber \\
&+\frac{1}{2\pi}\int_{0}^{\infty}d\xi\,  \frac{\mathrm{Tr}\left[\mathbf{d}\mathbf{d}\cdot \partial_{z}\underline{G}(\mathbf{r}_{a},\mathbf{r}, \imath \xi)_{\vert \mathbf{r}=\mathbf{r}_{a}}\right]}{\tilde{\omega}_{a}(\mathbf{r}_{a})+(-\imath \xi+\imath\gamma(\mathbf{r}_{a}))} .
\nonumber
\end{align}
\end{subequations}
Equation \eqref{imgQRT} is then recovered by using the definition in Eq.\eqref{alpha2ndOrder}.

It is also interesting to show that the non-Markovian contribution in Eq. \eqref{exactC} is 
directly responsible for the difference between the result of the QRT in Eq. \eqref{imgQRT} and the 
equation we obtain via the FDT in Eq. \eqref{imgFDT}. Upon inserting the second term on the 
right-hand side of Eq. \eqref{exactC} into the expression of the Casimir-Polder force (see 
Eq. \eqref{source1}), we have
\begin{align}
F^{\rm nM}_{\rm CP}&=
\mathrm{Re}\bigg(\frac{2}{\pi}\int_{0}^{\infty}d\omega\, \int_{0}^{\infty}d\tau e^{-\imath \omega \tau}\nonumber\\
&\hspace{1.0cm}\mathrm{Tr}\bigg[\left\{\frac{\hbar}{\pi} \!\int_{0}^{\infty} \!\!d\xi  
e^{-\xi\tau}[\underline{\alpha}_{I}(\omega)]_{\vert\omega=-\imath\xi+0^{+}}\right\}\nonumber\\
&\hspace{3.0cm}\cdot \partial_{z}\underline{G}_{I}(\mathbf{r}_{a},\mathbf{r},\omega)_{\vert \mathbf{r}=\mathbf{r}_{a}}\bigg]\bigg)\nonumber\\
&=
\mathrm{Tr}\bigg[\frac{\hbar}{\pi} \!\int_{0}^{\infty} \!\!d\xi \frac{ \underline{\alpha}(\imath\xi)-\underline{\alpha}(-\imath\xi)}{2} \nonumber\\
& \hspace{1.5cm}\cdot \frac{2}{\pi}\int_{0}^{\infty}d\omega\,
\frac{\omega\partial_{z}\underline{G}_{I}(\mathbf{r}_{a},\mathbf{r},\omega)_{\vert \mathbf{r}=\mathbf{r}_{a}}}{\omega^{2}+\xi^{2}}\bigg]
\nonumber\\
&=
\frac{\hbar}{\pi} \!\int_{0}^{\infty} \!\!d\xi \mathrm{Tr}\bigg[\frac{ \underline{\alpha}(\imath\xi)-\underline{\alpha}(-\imath\xi)}{2} \nonumber\\
& \hspace{3.0cm} \cdot 
\partial_{z}\underline{G}(\mathbf{r}_{a},\mathbf{r},\imath \xi)_{\vert \mathbf{r}=\mathbf{r}_{a}}\bigg] ,
\end{align} 
where we have already assumed that the polarization tensor is symmetric (see Eq.\eqref{alpha2ndOrder}). In
addition, in the last step we have used the Kramer-Kronig relation 
\cite{Landau80} 
(for simplicity, we have suppressed the dependence of the polarizability on $\mathbf{r}_{a}$). When added 
to Eq. \eqref{imgQRT} the above expression reproduces Eq. \eqref{imgFDT}.


\section{Non-Markovian contribution to quantum friction}
\label{FricNM}

The importance of  non-Markovianity to quantum friction can be assessed by considering the term 
$\underline{S}_{\rm nM}(\mathbf{k}\cdot\mathbf{v}-\omega)$ defined in Eq. \eqref{quantity}.
Using Eqs. \eqref{friction2} and \eqref{quantity} we obtain (for simplicity we consider 
$\mathbf{v}=v_{x}\mathbf{x}$)

\begin{align}
F^{\rm nM}_{\rm fric}&=
\frac{2\hbar}{\pi}\int\frac{d^{2}\mathbf{k}}{(2\pi)^{2}}k_{x}\int_{k_{x}v_{x}}^{\infty}d\omega\nonumber\\  
&\times\mathrm{Tr}\left[
\mathrm{Re}\sum_{i}\left(
\frac{\imath\mathrm{ Res[\underline{\alpha}(\Omega_{i})]}}{\Omega_{i}+\omega-k_{x}v_{x}}\right)\cdot \underline{G}^{\rm s}_{I}(\mathbf{k},z_{a}, \omega)\right]\nonumber\\
&+\frac{2\hbar}{\pi}\int\frac{d^{2}\mathbf{k}}{(2\pi)^{2}}k_{x}\int_{0}^{k_{x}v_{x}}d\omega  \nonumber\\
&\times\mathrm{Tr}\left[
\mathrm{Re}\sum_{i}\left(
\frac{\imath\mathrm{ Res[\underline{\alpha}(\Omega_{i})]}}{\Omega_{i}+k_{x}v_{x}-\omega}\right)\cdot \underline{G}^{\rm s}_{I}(\mathbf{k},z_{a}, \omega)\right]
\end{align} 
Recalling that $ \underline{G}^{\rm s}_{I}(\mathbf{k},z_{a}, \omega=0)=0$, we have that the leading 
term in the expansion for low velocities is provided by the first term on the right-hand side of this
equation. Consequently, we obtain in leading order
\begin{align}
F^{\rm nM}_{\rm fric}
&\approx
\frac{4\hbar v_{x}}{\pi}\int\frac{d^{2}\mathbf{k}}{(2\pi)^{2}}k_{x}^{2}\int_{0}^{\infty}d\omega \nonumber\\  
&\times\mathrm{Tr}\left[
\mathrm{Re}\sum_{i}\left(
\frac{\imath\mathrm{ Res[\underline{\alpha}(\Omega_{i})]}}{(\Omega_{i}+\omega)^{2}}\right)\cdot \underline{G}^{\rm s}_{I}(\mathbf{k},z_{a}, \omega)\right] .
\end{align} 
This exactly compensates 
the contribution arising from the QRT (cfr. Eq. \eqref{frictionQRT}). 



\begin{thebibliography}{10}

\bibitem{Casimir48a}
H.~B.~G. Casimir and D. Polder, {\em The Influence of Retardation on the
  London-van der Waals Forces}, Phys. Rev. {\bf 73},  360  (1948).

\bibitem{Pendry97}
J.~B. Pendry, {\em Shearing the vacuum - quantum friction}, J. Phys.: Condes.
  Matter {\bf 9},  10301  (1997).

\bibitem{Volokitin07}
A.~I. Volokitin and B.~N.~J. Persson, {\em Near-field radiative heat transfer
  and noncontact friction}, Rev. Mod. Phys. {\bf 79},  1291  (2007).

\bibitem{Gardiner91}
C. Gardiner, {\em Quantum Noise} (Springer-Verlag, Berlin, 1991).

\bibitem{Breuer02}
H. Breuer and F. Petruccione, {\em The Theory of Open Quantum Systems} (Oxford
  University Press, Oxford, 2002).

\bibitem{John90}
S. John and J. Wang, {\em Quantum electrodynamics near a photonic band gap:
  Photon bound states and dressed atoms}, Phys. Rev. Lett. {\bf 64},  2418
  (1990).

\bibitem{Vats02}
N. Vats, S. John, and K. Busch, {\em Theory of fluorescence in photonic
  crystals}, Phys. Rev. A {\bf 65},  043808  (2002).

\bibitem{Hoeppe12}
U. Hoeppe, C. Wolff, J. K{\"u}chenmeister, J. Niegemann, M. Drescher, H.
  Benner, and K. Busch, {\em Direct Observation of Non-Markovian Radiation
  Dynamics in 3D Bulk Photonic Crystals}, Phys. Rev. Lett. {\bf 108},  043603
  (2012).

\bibitem{Intravaia11}
F. Intravaia, C. Henkel, and M. Antezza,  in {\em Casimir Physics}, Vol.~834 of
  {\em Lecture Notes in Physics}, edited by D. Dalvit, P. Milonni, D. Roberts,
  and F. da~Rosa (Springer, Berlin / Heidelberg, 2011), pp.\ 345--391.

\bibitem{Buhmann04}
S.~Y. Buhmann, L. Knoell, D.-G. Welsch, and H.~T. Dung, {\em Casimir-Polder
  forces: A nonperturbative approach}, Phys. Rev. A {\bf 70},  052117  (2004).

\bibitem{Dzyaloshinskii61}
I.~E. Dzyaloshinskii, E.~M. Lifshitz, and L.~P. Pitaevskii, {\em General Theory
  of the van der Waals' Forces}, Sov. Phys. Usp. {\bf 4},  153  (1961).

\bibitem{Scheel09}
S. Scheel and S.~Y. Buhmann, {\em Casimir-Polder forces on moving atoms}, Phys.
  Rev. A {\bf 80},  042902  (2009).

\bibitem{Barton10b}
G. Barton, {\em On van der Waals friction. II: Between atom and half-space},
  New J. Phys. {\bf 12},  113045  (2010).

\bibitem{Intravaia14}
F. Intravaia, R.~O. Behunin, and D.~A.~R. Dalvit, {\em Quantum friction and
  fluctuation theorems}, Phys. Rev. A {\bf 89},  050101(R)  (2014).

\bibitem{Intravaia15}
F. Intravaia, V.~E. Mkrtchian, S.~Y. Buhmann, S. Scheel, D.~A.~R. Dalvit, and
  C. Henkel, {\em Friction forces on atoms after acceleration}, J. Phys.:
  Condes. Matter {\bf 27},  214020  (2015).

\bibitem{Huttner92}
B. Huttner and S.~M. Barnett, {\em Quantization of the electromagnetic field in
  dielectrics}, Phys. Rev. A {\bf 46},  4306  (1992).

\bibitem{Rosa10}
F.~S.~S. Rosa, D.~A.~R. Dalvit, and P.~W. Milonni, {\em Electromagnetic energy,
  absorption, and Casimir forces: Uniform dielectric media in thermal
  equilibrium}, Phys. Rev. A {\bf 81},  033812  (2010).

\bibitem{Rosa11}
F.~S.~S. Rosa, D.~A.~R. Dalvit, and P.~W. Milonni, {\em Electromagnetic energy,
  absorption, and Casimir forces. II. Inhomogeneous dielectric media}, Phys.
  Rev. A {\bf 84},  053813  (2011).

\bibitem{Intravaia12b}
F. Intravaia and R. Behunin, {\em Casimir effect as a sum over modes in
  dissipative systems}, Phys. Rev. A {\bf 86},  062517  (2012).

\bibitem{Milonni73}
P.~W. Milonni, J.~R. Ackerhalt, and W.~A. Smith, {\em Interpretation of
  Radiative Corrections in Spontaneous Emission}, Phys. Rev. Lett. {\bf 31},
  958  (1973).

\bibitem{Dalibard82}
J. Dalibard, J. Dupont-Roc, and C. Cohen-Tannoudji, {\em Vacuum fluctuations
  and radiation reaction : identification of their respective contributions},
  J. Phys. France {\bf 43},  1617  (1982).

\bibitem{Ford06}
G.~W. Ford and R.~F. O'Connell, {\em A Quantum Violation of the Second Law?},
  Phys. Rev. Lett. {\bf 96},  020402  (2006).

\bibitem{Intravaia08}
F. Intravaia and C. Henkel, {\em Casimir energy and entropy between dissipative
  mirrors}, J. Phys. A: Math. Gen. {\bf 41},  164018 (9pp)  (2008).

\bibitem{Compagno05}
G. Compagno, R. Passante, and F. Persico, {\em Atom-field interactions and
  dressed atoms} (Cambridge University Press, Cambridge, UK, 2005), Vol.~17.

\bibitem{Kubo57}
R. Kubo, {\em Statistical-Mechanical Theory of Irreversible Processes. I.
  General Theory and Simple Applications to Magnetic and Conduction Problems},
  J. Phys. Soc. Jap. {\bf 12},  570  (1957).

\bibitem{Kubo66}
R. Kubo, {\em The fluctuation-dissipation theorem}, Rep. Prog. Phys. {\bf 29},
  255  (1966).

\bibitem{Haag67}
R. Haag, N. Hugenholtz, and M. Winnink, {\em On the equilibrium states in
  quantum statistical mechanics}, Comm. Math. Phys. {\bf 5},  215  (1967).

\bibitem{Callen51}
H.~B. Callen and T.~A. Welton, {\em Irreversibility and Generalized Noise},
  Phys. Rev. {\bf 83},  34  (1951).

\bibitem{Wylie84}
J.~M. Wylie and J.~E. Sipe, {\em Quantum electrodynamics near an interface},
  Phys. Rev. A {\bf 30},  1185  (1984).

\bibitem{McLachlan63}
A.~D. McLachlan, {\em Retarded Dispersion Forces in Dielectrics at Finite
  Temperatures}, Proc. R. Soc. Lond. A {\bf 274},  80  (1963).

\bibitem{Lax63}
M. Lax, {\em Formal Theory of Quantum Fluctuations from a Driven State}, Phys.
  Rev. {\bf 129},  2342  (1963).

\bibitem{Mandel95}
L. Mandel and E. Wolf, {\em Optical Coherence and Quantum Optics} (Cambridge
  University Press, New York, 1995).

\bibitem{Onsager31}
L. Onsager, {\em Reciprocal Relations in Irreversible Processes. II.}, Phys.
  Rev. {\bf 38},  2265  (1931).

\bibitem{Onsager31a}
L. Onsager, {\em Reciprocal Relations in Irreversible Processes. I.}, Phys.
  Rev. {\bf 37},  405  (1931).

\bibitem{Talkner86}
P. Talkner, {\em The failure of the quantum regression hypothesis}, Ann. Phys.
  {\bf 167},  390  (1986).

\bibitem{Ford96a}
G.~W. Ford and R.~F. O'Connell, {\em There is No Quantum Regression Theorem},
  Phys. Rev. Lett. {\bf 77},  798  (1996).

\bibitem{Ford00}
G. Ford and R. O'Connell, {\em Driven systems and the Lax formula}, Opt. Comm.
  {\bf 179},  451   (2000).

\bibitem{Lax00}
M. Lax, {\em The Lax-Onsager regression `theorem' revisited}, Opt. Comm. {\bf
  179},  463   (2000).

\bibitem{Ford00a}
G. Ford and R. O'Connell, {\em Comment on The Lax-Onsager Regression Theorem
  revisited}, Opt. Comm. {\bf 179},  477   (2000).

\bibitem{Note1}
It is important to point out, however, that this is only an approximation valid
  in weak coupling. In reality, it is the global system that goes to a highly
  correlated stationary equilibrium state; in general, the reduced state of the
  atom is given by a mixture of ground and excited states \cite
  {Intravaia08,Klich12}.

\bibitem{Milonni04}
P.~W. Milonni and R.~W. Boyd, {\em Influence of radiative damping on the
  optical-frequency susceptibility}, Phys. Rev. A {\bf 69},  023814  (2004).

\bibitem{Lach12}
G. Lach, M. DeKieviet, and U.~D. Jentschura, {\em Enhancement of Blackbody
  Friction due to the Finite Lifetime of Atomic Levels}, Phys. Rev. Lett. {\bf
  108},  043005  (2012).

\bibitem{Jentschura15}
D. Jentschura, U., G. Lach, M. De~Kieviet, and K. Pachucki, {\em One-Loop
  Dominance in the Imaginary Part of the Polarizability: Application to
  Blackbody and Noncontact van der Waals Friction}, Phys. Rev. Lett. {\bf 114},
   043001  (2015).

\bibitem{Buhmann12}
S.~Y. Buhmann, S. Scheel, S.~{\AA}. Ellingsen, K. Hornberger, and A. Jacob,
  {\em Casimir-Polder interaction of fullerene molecules with surfaces}, Phys.
  Rev. A {\bf 85},  042513  (2012).

\bibitem{Steck08}
D.~A. Steck, Technical report, Oregon Center for Optics and Department of
  Physics, University of Oregon (unpublished).

\bibitem{Gullo14}
N.~L. Gullo, I. Sinayskiy, T. Busch, and F. Petruccione, {\em Non-Markovianity
  criteria for open system dynamics}, arXiv preprint arXiv:1401.1126  (2014).

\bibitem{Ali15}
M.~M. Ali, P.-Y. Lo, M.~W.-Y. Tu, and W.-M. Zhang, {\em Non-Markovianity
  measure using two-time correlation functions}, Phys. Rev. A {\bf 92},  062306
   (2015).

\bibitem{Knight76}
P.~L. Knight and P.~W. Milonni, {\em Long-time deviations from exponential
  decay in atomic spontaneous emission theory}, Phys. Lett. A {\bf 56},  275
  (1976).

\bibitem{Berman10}
P.~R. Berman and G.~W. Ford,  in {\em Advances In Atomic, Molecular, and
  Optical Physics}, edited by E. Arimondo, P.~R. Berman, and C.~C. Lin
  (Academic Press, Amsterdam, 2010), Vol.~59, p.\ 175.

\bibitem{Davidson70}
R. Davidson and J.~J. Kozak, {\em On the Relaxation to Quantum-Statistical
  Equilibrium of the Wigner-Weisskopf Atom in a One-Dimensional Radiation
  Field. I. A Study of Spontaneous Emission}, J. Math. Phys. {\bf 11},  189
  (1970).

\bibitem{Davidson71}
R. Davidson and J.~J. Kozak, {\em Relaxation to Quantum-Statistical Equilibrium
  of the Wigner-Weisskopf Atom in a One-Dimensional Radiation Field. III. The
  Quantum-Mechanical Solution}, J. Math. Phys. {\bf 12},  903  (1971).

\bibitem{Wodkiewicz76}
K. W{\'o}dkiewicz and J.~H. Eberly, {\em Markovian and non-markovian behavior
  in two-level atom fluorescence}, Ann. Phys. {\bf 101},  574  (1976).

\bibitem{Weisskopf30}
V. Weisskopf and E. Wigner, {\em Berechnung der nat{\"u}rlichen Linienbreite
  auf Grund der Diracschen Lichttheorie}, Z. Physik {\bf 63},  54  (1930).

\bibitem{Foerster72}
T. von Foerster, {\em Quantum Theory of a Damped Two-Level Atom}, Am. J. Phys.
  {\bf 40},  854  (1972).

\bibitem{Cohen-Tannoudji98}
C. Cohen-Tannoudji, J. Dupont-Roc, and G. Grynberg, {\em Atom-photon
  interactions} (John Wiley and Sons Inc., New York, 1998).

\bibitem{Dedkov02a}
G. Dedkov and A. Kyasov, {\em Electromagnetic and fluctuation-electromagnetic
  forces of interaction of moving particles and nanoprobes with surfaces: A
  nonrelativistic consideration}, Phys. Solid State {\bf 44},  1809  (2002).

\bibitem{Hoye14}
J.~S. H{\o}ye and I. Brevik, {\em Casimir friction at zero and finite
  temperatures}, Eur. Phys. J. D {\bf 68},  1  (2014).

\bibitem{Maghrebi12}
M.~F. Maghrebi, R.~L. Jaffe, and M. Kardar, {\em Spontaneous Emission by
  Rotating Objects: A Scattering Approach}, Phys. Rev. Lett. {\bf 108},  230403
   (2012).

\bibitem{Maghrebi13a}
M.~F. Maghrebi, R. Golestanian, and M. Kardar, {\em Quantum Cherenkov radiation
  and noncontact friction}, Phys. Rev. A {\bf 88},  042509  (2013).

\bibitem{Polder71}
D. Polder and M. Van~Hove, {\em Theory of Radiative Heat Transfer between
  Closely Spaced Bodies}, Phys. Rev. B {\bf 4},  3303  (1971).

\bibitem{Obrecht07}
J.~M. Obrecht, R.~J. Wild, M. Antezza, L.~P. Pitaevskii, S. Stringari, and
  E.~A. Cornell, {\em Measurement of the Temperature Dependence of the
  Casimir-Polder Force}, Phys. Rev. Lett. {\bf 98},  063201  (2007).

\bibitem{Intravaia16a}
F. Intravaia, R.~O. Behunin, C. Henkel, K. Busch, and D.~A.~R. Dalvit, to be
  submitted.

\bibitem{Frolov86}
V. Frolov and V. Ginzburg, {\em Excitation and radiation of an accelerated
  detector and anomalous doppler effect}, Phys. Lett. A {\bf 116},  423
  (1986).

\bibitem{Ginzburg96}
V.~L. Ginzburg, {\em Radiation by uniformly moving sources (Vavilov--Cherenkov
  effect, transition radiation, and other phenomena)}, Physics-Uspekhi {\bf
  39},  973  (1996).

\bibitem{Cerenkov37}
P.~A. {\v C}erenkov, {\em Visible Radiation Produced by Electrons Moving in a
  Medium with Velocities Exceeding that of Light}, Phys. Rev. {\bf 52},  378
  (1937).

\bibitem{Frank37}
I. Frank and I. Tamm, {\em Coherent visible radiation of fast electrons passing
  through matter}, C.R. Acad. Sci. URSS {\bf 14},  109  (1937).

\bibitem{Maslovski11b}
S.~I. Maslovski and M.~G. Silveirinha, {\em Casimir forces at the threshold of
  the Cherenkov effect}, Phys. Rev. A {\bf 84},  062509  (2011).

\bibitem{Pieplow15}
G. Pieplow and C. Henkel, {\em Cherenkov friction on a neutral particle moving
  parallel to a dielectric}, J. Phys.: Condes. Matter {\bf 27},  214001
  (2015).

\bibitem{Klatt16}
J. Klatt, R. Bennett, and S.~Y. Buhmann, eprint: arXiv:1601.02765.

\bibitem{Dedkov03}
G. Dedkov and A. Kyasov, {\em The relativistic theory of fluctuation
  electromagnetic interactions of moving neutral particles with a flat
  surface}, Phys. Solid State {\bf 45},  1815  (2003).

\bibitem{Volokitin08}
A.~I. Volokitin and B.~N.~J. Persson, {\em Theory of the interaction forces and
  the radiative heat transfer between moving bodies}, Phys. Rev. B {\bf 78},
  155437  (2008).

\bibitem{Silveirinha14a}
M.~G. Silveirinha, {\em Optical Instabilities and Spontaneous Light Emission by
  Polarizable Moving Matter}, Phys. Rev. X {\bf 4},  031013  (2014).

\bibitem{Jackson75}
J. Jackson, {\em Classical Electrodynamics} (John Wiley and Sons Inc., New
  York, 1975).

\bibitem{Lambrecht06}
A. Lambrecht, P.~A.~M. Neto, and S. Reynaud, {\em The Casimir effect within
  scattering theory}, New J. Phys. {\bf 8},  243  (2006).

\bibitem{Messina09}
R. Messina, D.~A.~R. Dalvit, P.~A.~M. Neto, A. Lambrecht, and S. Reynaud, {\em
  Dispersive interactions between atoms and nonplanar surfaces}, Phys. Rev. A
  {\bf 80},  022119  (2009).

\bibitem{Landau80}
L. Landau, E. Lifshitz, and L. Pitaevskii, {\em Course of Theoretical Physics:
  Electrodinamics in Continuous Media}, 2nd ed. (Butterworth-Heinmann, Oxford,
  1980).

\bibitem{Klich12}
I. Klich, {\em Entanglement of a Quantum Field with a Dispersive Medium}, Phys.
  Rev. Lett. {\bf 109},  061601  (2012).

\end{thebibliography}

\end{document}